\documentclass[a4paper,12pt]{article}
\pdfoutput=1

\makeatletter

\@addtoreset{equation}{section}
\makeatother
\usepackage{tikz-cd}
\usepackage{tikz-3dplot}

\usepackage{caption}
\usepackage{epigraph}

\usetikzlibrary{decorations.pathmorphing}

\usetikzlibrary{arrows.meta}

\usetikzlibrary{decorations.markings}
\usepackage{amsfonts}
\usepackage{amsmath}
\usepackage{latexsym}

\usepackage{amsthm}

\usepackage{mathrsfs}

\newcommand{\R}{\mathbb{R}}

\newcommand{\C}{\mathbb{C}}

\newcommand{\ev}{\mathrm{ev}}

\def\less{\ll}

\def\be{\begin{equation}}
\def\ee{\end{equation}}
\def\bea{\begin{eqnarray}}
\def\eea{\end{eqnarray}}
\def\({\left(}
\def\){\right)}
\def\<{\left<}
\def\>{\right>}

\def\[{\left[}
\def\]{\right]}

\def\be{\begin{equation}}
\def\ee{\end{equation}}
\def\bea{\begin{eqnarray}}
\def\eea{\end{eqnarray}}

\def\({\left(}
\def\){\right)}
\def\<{\left<}
\def\>{\right>}

\def\tr{{\mbox{tr}}}
\def\be{\begin{equation}}
\def\ee{\end{equation}}
\def\bea{\begin{eqnarray}}
\def\eea{\end{eqnarray}}
\def\ben{\begin{eqnarray*}}
\def\een{\end{eqnarray*}}
\def\({\left(}
\def\){\right)}
\def\<{\left<}
\def\>{\right>}
\def\!{\right|}
\def\|{\left|}

\def\[{\left[}
\def\]{\right]}

\def\+{\bar}
\def\mb{\mathbb}
\def\tr{{\mbox{tr}}}

\def\L{{\cal{L}}}
\def\t{\widetilde}
\def\A{{\cal{A}}}
\def\B{{\cal{B}}}
\def\C{{\cal{C}}}
\def\R{{\cal{R}}}

\def\J{{\cal{J}}}
\def\N{{\cal{N}}}

\def\F{{\cal{F}}}

\def\O{{\cal{O}}}

\def\P{{\cal{P}}}
\def\L{{\cal{L}}}

\def\J{{\overline{\L}}}

\def\L{{\cal{L}}}

\def\eps{{\cal{\varepsilon}}}

\def\E{{\cal{E}}}

\def\F{{\cal{F}}}

\def\V{{\cal{V}}}

\def\U{{\cal{U}}}

\def\h{\widehat}

\def\P{\mathscr{P}}
\def\N{\mathscr{N}}

\def\*{{\times}}

\begin{document}

\setlength{\unitlength}{1mm}

\pagestyle{empty}
\vskip-10pt
\vskip-10pt
\hfill 
\begin{center}
\vskip 3truecm
{\Large \bf
The tensor multiplet in loop space}
\vskip 2truecm
{\large \bf
Dongsu Bak, Andreas Gustavsson}
\vspace{1cm} 
\begin{center} 
Physics Department, University of Seoul, Seoul 02504 KOREA
\end{center}
\end{center}
\vskip 2truecm
{\abstract We reformulate the abelian tensor multiplet on a curved spacetime with at least two supercharges in a cohomological form where all the bosonic and fermionic fields become tensor fields. These tensor fields are rewritten as fields in loop space by a transgression map. There are two lightlike conformal Killing vectors. By decomposing the spacetime tensor fields in transverse and parallel components to these Killing vectors, we obtain the equations of motion in loop space by closing the supersymmetry variations on-shell. We generalize to nonabelian gauge groups. By closing supersymmetry variations we obtain nonabelian fermionic equations of motion in loop space.}

\vfill
\vskip4pt
\eject
\pagestyle{plain}

\section{Introduction}
The abelian tensor multiplet is difficult to generalize to nonabelian gauge groups because of the two-form gauge potential $B_{MN}$. There is a no-go theorem that says that no reparametrization invariant Wilson surface can be constructed out of a nonabelian two-form gauge potential \cite{Teitelboim:1985ya}, \cite{Teitelboim:1972vw} alone. However, in loop space the two-form gauge potential becomes a one-form through a transgression map from spacetime to loop space \cite{Hofman:2002ey}, which is an integration of the two-form around a loop $C$,
\bea
\A(C) &=& \int_C B_{MN}(C(s)) \delta C^M(s) \dot{C}^N(s) ds\label{t1}
\eea
Since this is now a one-form gauge field, it seems, at least naively, that it should have a generalization to nonabelian gauge groups, despite the above mentioned no-go theorem. Here the integral is around a loop $C$ parametrized and embedded in spacetime as $s\mapsto C^M(s)$. Its tangential vector is denoted as $\dot{C}^M(s)$ and $\delta C^M(s)$ constitute a basis for the cotangent bundle over loop space that are simply wedge multiplied together as $\delta C^M(s) \wedge \delta C^N(s)$ where the wedge product in loop space is directly inherited from the wedge product in spacetime. We will refer to $\A(C)$ as a loop space field. All loop space fields will be invariant under reparametrizations of the loop. They will also all be Weyl invariant. Our loop space is the set of all loops embedded in spacetime and the loops are not required to have a common base point, which means that we have a free loop space. Reparametrization invariance means that any two loops that differ only by their parametrization shall be identified as the same loop and we will assume that the loops are everywhere smooth for simplicity. By allowing the loops to have cusp and self-intersection points we would likely discover new  interesting physical effects, which deserves a separate treatment.

We may construct a nonabelian gauge potential in loop space as follows,\bea
\A(C) &=& \int B_{MN}^a(C(s)) \delta C^M(s) n^N(C(s)) t_a(s) ds\label{t2}
\eea
Here $B_{MN}^a$ is a nonabelian two-form in component form with $a$ being a gauge group index in the adjoint representation, $n^M$ is a vector field, and as it turns out, this will be a conformal Killing vector field on the spacetime manifold and $t_a(s)$ are generators of a loop algebra
\bea
[t_a(s),t_b(s')] &=& i f_{ab}{}^c \delta(s-s') t_c(s)
\eea
where $f_{ab}{}^c$ are structure constants of the gauge group. The resulting nonabelian gauge field $\A$ now has some nice properties. Firstly, because of the delta function in the loop algebra, all commutators such as $\A \wedge \A$ will be local expressions integrated around the loop.\footnote{For a different approach that also results in a local $\A \wedge \A$ but which does not use a loop algebra but instead a vanishing fake curvature condition, see \cite{Schreiber:2005ff}, \cite{Baez:2004in}, \cite{Girelli:2003ev}.} Secondly, $\A$ is reparametrization invariant as a consequence of that the generators $t_a(s)$ transform as one-forms. As we will show, this construction is consistent only if we perform dimensional reduction along $n^M$. Nevertheless, this reformulation may have some advantages compared to the standard formulation of five-dimensional super Yang-Mills theory. For example, it may be better suited for defining the nonabelian Wilson surface that in loop space would be defined as a Wilson line
\bea
P \exp \(i e \int_{\Gamma} \A\)
\eea
where $P$ denotes path-ordering along a path $\Gamma$ in loop space, which is a surface in spacetime that is foliated by loops in a certain way. While we have reasons to believe that this definition will be independent of the way that the surface is foliated by loops thanks to the loop algebra \cite{Teitelboim:1985ya}, \cite{Teitelboim:1972vw}, \cite{Gustavsson:2005fp}, we plan to address this question in detail in a future publication.

The structure of this paper is as follows. In section \ref{tva} we start with the abelian $(1,0)$ tensor multiplet with two chiral spinors $\lambda_I$ for $I = 1,2$. We pick two commuting supersymmetry parameters $\eps_I$ which are conformal Killing spinors and use them to map the two spinors into tensor fields. We then recall a basic geometrical fact from \cite{Bak:2024ihe} that having two commuting conformal Killing spinors implies that the Lorentzian six-manifold possesses two commuting lightlike conformal Killing vector fields $U$ and $V$. These two vector fields enable us to decompose any tensor field into transverse and longitudinal components where the transverse components are orthogonal to both $U$ and $V$. In section \ref{tre} we introduce a transgression map from spacetime fields to loop space fields, essentially by integration over the loop as in (\ref{t1}), but, as we will see, without integrating. We obtain Bianchi identities, supersymmetry variations and matter field equations of motion in loop space in sections \ref{fyra}, \ref{fem} and \ref{sex}. We then repeat the same steps for a nonabelian gauge group. We propose a nonabelian transgression map (\ref{t2}), obtain Bianchi identities, supersymmetry variations and matter field equations of motion in sections \ref{sju}, \ref{atta} and \ref{nio}. We show that our nonabelian equations of motion are supersymmetric and that they reduce to equations of motion in five-dimensional super Yang-Mills theory in sections \ref{nio} and \ref{tio}. In section \ref{elva} we discuss our results.

\section{The abelian tensor multiplet}\label{tva}
In this section we present the abelian tensor multiplet in six dimensions in the cohomological form \cite{Bak:2024ihe}. This is the curved space generalization and uplift to six dimensions of the corresponding five-dimensional vector multiplet in the cohomological form \cite{Kallen:2012va}. The cohomological form was first used in a localization computation in 2d Yang-Mills theory by Witten \cite{Witten:1992xu} for the BRST symmetry. Later a cohomological form of 4d super Yang-Mills was used by Pestun \cite{Pestun:2007rz} for supersymmetric localization and this technique has subsequently been applied to other supersymmetric theories including five dimensional super Yang-Mills theory \cite{Kallen:2012va}. We will use a similar kind of cohomological form of the tensor multiplet in six dimensions. Our motivation is not to use it for localization but to transgress the fermionic field into loop space. In the cohomological form the spinor fields become a set of tensor fields that we can transgress to loop space by integrating them over the loop. 

The $(1,0)$ tensor multiplet has a gauge potential $B_{MN}$, one real scalar field $\phi$ and two symplectic-Majorana chiral fermions $\lambda_I$ where $I =1,2$. The supersymmetry variations are
\bea
\delta \phi &=& - i \bar\eps^I \lambda_I\cr
\delta B_{MN} &=& i \bar\eps^I \Gamma_{MN} \lambda_I\cr
\delta \lambda_I &=& \frac{1}{12} \Gamma^{MNP} \eps_I H_{MNP} + \Gamma^M \eps_I \partial_M \phi + 4 \eta_I \phi
\eea
where
\bea
H_{MNP} &=& \partial_M B_{NP} + \partial_N B_{PM} + \partial_P B_{MN}
\eea
Here the supersymmetry parameters satisfy the conformal Killing spinor equation
\bea
\nabla_M \eps_I &=& \Gamma_M \eta_I
\eea
Our spinor conventions are summarized in appendix \ref{spinors}. There can be various supersymmetric actions that one may consider for a selfdual tensor field. But the most straightforward way is to use the Maxwell action for a non-selfdual tensor field, which can be supersymmetrized. This action is given by
\bea
S &=& \int d^6 x \sqrt{-g} \L\cr
\L &=& - \frac{1}{24} H_{MNP}^2 - \frac{1}{2} (\nabla_M \phi)^2 + \frac{i}{2} \bar\lambda^I \Gamma^M \nabla_M \lambda_I - \frac{R}{10} \phi^2
\eea
Here $g_{MN}$ is a Lorentzian metric and $R$ is the Ricci curvature scalar.
The selfduality of $H_{MNP}$ is an equation of motion that is required for on-shell closure of the supersymmetry variations but which we can not derive from this action. 

The two $\eps_I$ for $I = 1,2$ may be taken to be either anticommuting or commuting. We will take them to be commuting and the supersymmetry variation $\delta$ is anticommuting. 

From these two $\eps_I$ one can show \cite{Bak:2024ihe} that there has to exist two commuting lightlike conformal Killing vectors $V_M$ and $U_M$ where
\bea
V_M &=& \bar\eps^I \Gamma_M \eps_I
\eea
That they are commuting means that their Lie bracket vanishes
\bea
[V,U]^M = V^N \partial_N U^M - U^N \partial_N V^M = 0
\eea
From the identity $\L_{[U,V]} = [\L_U,\L_V]$ it then follows that the Lie derivatives commute
\bea
[\L_U,\L_V] &=& 0
\eea
We define a (possibly spacetime dependent) normalization factor 
\bea
\N &=& g_{MN} U^M V^N
\eea
and we define their Weyl invariant curvatures
\bea
V_{MN} &=& \partial_M \(\frac{V_N}{\N}\) - \partial_N \(\frac{V_M}{\N}\)\cr
U_{MN} &=& \partial_M \(\frac{U_N}{\N}\) - \partial_N \(\frac{U_M}{\N}\)
\eea
In appendix \ref{UoV} we show that these are transverse to both $U$ and $V$ and selfdual and antiselfdual in this transverse space, respectively. 

We define fermionic tensor fields as
\bea
\Psi_{MN} &=& \bar\eps^I \Gamma_{MN} \lambda_I\cr
\Psi_N &=& U^M \Psi_{MN}\cr
\psi &=& \frac{1}{\N} V^M \Psi_M
\eea
and we now see that it is essential to take $\eps_I$ to be commuting in order for these fermionic tensor fields to be anticommuting. 

From the above definitions it follows that 
\bea
\Psi_{MN} V^N &=& V_M \psi
\eea
This relation follows because of the property
\bea
\Gamma_M \eps_I V^M &=& 0
\eea
that one can prove by using a Fierz identity. 

The supersymmetry variation expressed in terms of these fermionic tensor fields reads
\bea
\delta \phi &=& - i \psi\cr
\delta B_{MN} &=& i \Psi_{MN}\cr
\delta \Psi_{MN} &=& - V_{MN} \N \phi + \frac{V_M}{\N} \partial_N \(\N\phi\) - \frac{V_N}{\N} \partial_M\(\N\phi\)\cr
&& - V^P H_{PMN}^+\cr
\delta \Psi_N &=& \partial_N \(\N\phi\) - \frac{V_N}{\N} \L_U \(\N\phi\)\cr
&& - V^P U^M H_{PMN}^+\cr
\delta \psi &=& \frac{1}{\N} \L_V \(\N\phi\)
\eea
We will decompose the tensor fields in transverse and longitudinal components to $U$ and $V$ as
\bea
\Psi_{MN} &=& \chi_{MN} + \frac{1}{\N} \(V_M \Psi_N - V_N \Psi_M\)\cr
\Psi_M &=& \psi_M + U_M \psi
\eea
We also define 
\bea
F_{MN} &=& \t{H_{MNP} V^P}\cr
G_{MN} &=& \t{H_{MNP} U^P}\cr
R_P &=& - V^M U^N H_{MNP}
\eea
Tilde indicates that trace parts have been removed, where we refer to a contraction with $U$ or $V$ as a trace. These traceless tensor fields will be also referred to as irreducible fields. A reducible tensor field, such as $\Psi_{MN}$ is expanded in irreducible components $\chi_{MN}$, $\psi_M$ and $\psi$. Similarly the reducible field strength is expanded in irreducible components as
\bea
H_{MNP} &=& h_{MNP} + \frac{3}{\N} U_{[M} F_{NP]} + \frac{3}{\N} V_{[M} G_{NP]} - \frac{6}{\N^2} U_{[M} V_N R_{P]}
\eea
The supersymmetry variations of these irreducible component fields are 
\bea
\delta \chi_{MN} &=& - V_{MN} \N \phi - F_{MN}^+\cr
\delta \psi_N &=& \t\partial_N \(\N\phi\) + R_N^+\cr
\delta F_{MN} &=& i \L_V \chi_{MN} + i V_{MN} \N \psi\cr
\delta G_{MN} &=& i \L_U \chi_{MN} + i U_{MN} \N \psi - i \t{\(\partial_M \psi_N - \partial_N \psi_M\)}\cr
\delta R_M &=& i \(\t\partial_M \(\N\psi\) - \L_V \psi_M\)
\eea
where
\bea
F_{MN}^{\pm} &=& \frac{1}{2} \(F_{MN} \pm \frac{1}{2} \E_{MNPQ} F^{PQ}\)\cr
R_M^{\pm} &=& \frac{1}{2} \(R_M \mp \frac{1}{6\N} \E_{MRST} h^{RST}\)
\eea
It is interesting to note that the supersymmetry variation of $F_{MN}$ is selfdual. 

In this formulation we have spontaneously broken Lorentz symmetry in Minkowski space since $V$ selects a preferred lightlike direction. Without breaking Lorentz symmetry, we can not extract $F_{MN}$ but must work with $H_{MNP}$ that shall be subject to the selfduality equation of motion
\bea
H_{MNP} &=& \frac{1}{6} \eps_{MNP}{}^{RST} H_{RST}
\eea
This selfduality induces the selfduality relations 
\bea
F_{MN} &=& \frac{1}{2} \E_{MN}{}^{PQ} F_{PQ}\cr
G_{MN} &=& - \frac{1}{2} \E_{MN}{}^{PQ} G_{PQ}\cr
R_Q &=& - \frac{\N}{6} \E_{Q}{}^{MNP} h_{MNP}\cr
h_{MNP} &=& \frac{1}{\N} \E_{MNP}{}^Q R_Q
\eea
where we define 
\bea
\E_{MNPQ} &=& \frac{1}{\N}\eps_{MNPQRS} V^R U^S 
\eea
with the property
\bea
\nabla_M \(\frac{1}{\N} \E^{MNPQ}\) &=& - \frac{3}{\N} \(U^{[NP} V^Q + V^{[NP} U^{Q]}\)
\eea
that we derive in eq (\ref{derivedE})) in the appendix. 

The breaking of Lorentz symmetry is in the formulation of the theory rather than a physical breaking, which corresponds to selecting two  supercharges out of eight in Minkowski space. That our formulation breaks some symmetries, like Lorentz symmetry, seems harmless in the abelian theory since we have an alternative formulation in terms of spinor fields where the Lorentz symmetry is manifest. 

The supersymmetric action is given by
\bea
S &=& \int d^6 x \sqrt{-g} \L\cr
\L &=& - \frac{1}{24} H_{MNP}^2 - \frac{1}{2\N^2} \(\nabla_M\(\N\phi\)\)^2\cr
&& - \frac{i}{4\N} \chi^{MN} \L_U \chi_{MN} + \frac{i}{\N} \chi^{MN} \partial_M \psi_N - \frac{i}{\N} V_{MN} \psi^M \psi^N\cr
&& + \frac{i}{2\N^2} \psi^M \L_V \psi_M - \frac{i}{\N^2} \psi^M \partial_M \(\N\psi\) - \frac{i}{\N^2} \psi \L_U \(\N\psi\)\label{spacetimeaction}
\eea
In appendix \ref{scalarR} we match the scalar field part of this action with the standard action for a conformally coupled scalar field in a curved background geometry. There are three fermionic equations of motion
\bea
\N^2 \nabla^M \(\frac{1}{\N^2}\psi_M\) + \frac{2}{\N} \L_U \(\N \psi\) &=& 0\label{firsteom}\\
\N^2 \t{\nabla^M\(\frac{1}{\N}\chi_{MN}\)} - \N V_{NP} \psi^P &=& \t{\partial}_N \(\N \psi\) - \L_V \psi_N\label{secondeom}\\
\L_U \chi_{MN} &=&  \t{\(\partial_M \psi_N - \partial_N \psi_M\)}^+\label{thirdeom}
\eea 
that are obtained from this action by varying $\psi$, $\psi_M$ and $\chi_{MN}$ respectively. Making a supersymmetry variation of the first equation (\ref{firsteom}) we obtain the scalar field equation of motion
\bea
\N^2 \nabla^M \(\frac{1}{\N^2} \partial_M \(\N\phi\)\) &=& 0
\eea
In the second fermionic equation of motion, (\ref{secondeom}), the traceless part is given by
\bea
\t{\nabla^M \(\frac{1}{\N} \chi_{MN}\)} &=& \nabla^M \(\frac{1}{\N} \chi_{MN}\) + \frac{1}{2\N} U_N V^{PQ} \chi_{PQ}
\eea
To verify this we first contract the right-hand side with $V^N$ and make an `integration by parts',
\bea
&& V^N \nabla^M \(\frac{1}{\N} \chi_{MN}\) + \frac{1}{2} V^{PQ} \chi_{PQ}\cr
&=& \nabla^M \(\frac{V^N}{\N} \chi_{MN}\) - \nabla^M \(\frac{V^N}{\N}\) \chi_{MN} + \frac{1}{2} V^{MN} \chi_{MN}\cr
&=& 0
\eea
In the last step we used the traceless property of $\chi_{MN}$ and the definition of $V_{MN}$. Next, if we contract instead with $U^N$ and repeat the same kind of computation, we get
\bea
- \frac{1}{2} U^{MN} \chi_{MN} &=& 0
\eea
This is zero because $U_{MN}$ is anti-selfdual and $\chi_{MN}$ is selfdual off-shell. Hence the right-hand side is traceless. 

The Bianchi identity is
\bea
\partial_{[M} H_{NPQ]} &=& 0
\eea
By various contractions with $U$ and $V$ one can show that this induces the following four Bianchi identities on the components,
\bea 
\E^{MNPQ} \partial_M h_{NPQ} &=& 0\cr
\L_V h_{MNP} &=& 3 \partial_{[M} F_{NP]}\cr
\L_U h_{MNP} &=& 3 \partial_{[M} G_{NP]}\cr
\L_V G_{MN} - \L_U F_{MN} &=& \t{\partial_M R_N - \partial_N R_M}
\eea
Here
\bea
\t{\partial_M R_N - \partial_N R_M} &=& \partial_M R_N - \partial_N R_M\cr
&& - \frac{U_M}{\N} \L_V R_N - \frac{V_M}{\N} \L_U R_N\cr
&& + \frac{U_N}{\N} \L_V R_M + \frac{V_N}{\N} \L_U R_M
\eea
is the traceless part. The small computation that is required to see this is 
\bea
V^M \(\partial_M R_N - \partial_N R_M\) &=& \L_V R_N
\eea
which uses the fact that $V^M R_M = 0$. A similar result holds for the contraction with $U^M$.

\section{The abelian transgression map}\label{tre}
We begin with reviewing the transgression map \cite{Brylinski}, \cite{BottTu}, \cite{Hofman:2002ey} that we will also refer to as the integrated transgression map because it is integrated over the loop. 

We then note that this integrated transgression map is not injective for one-forms that are mapped to zero-forms in loop space because any exact one-form is transgressed to a vanishing zero-form in loop space. Another problem that arises only for the case of a one-form being transgressed is the huge loss of information that does not occur for higher-rank spacetime forms when these are transgressed into loop space. These two considerations lead us to introduce an unintegrated transgression map, which enables us to probe better the local dependence around the loop of the loop space fields in a way that makes it more suitable for our construction of a field theory in loop space. To our knowledge the unintegrated transgression map has not appeared previously in the literature.

\subsection{Review of the integrated transgression map}
We let $M$ be a smooth six-dimensional spacetime manifold. We define free loop space $LM$ as the set of smooth maps from $S^1$ to $M$. 

The evaluation map $\text{ev}: S^1 \times LM \to M$ is a map that gives the spacetime point of the loop $C\in LM$ at the parameter value $s\in S^1$ as
\bea
(s,C) \mapsto C(s)
\eea
The transgression map is a map $\tau: \Omega^p(M) \to \Omega^{p-1}(LM)$ that is defined as follows. First we pull back the $p$-form $\omega \in \Omega^p(M)$ to $\Omega^p(S^1 \times LM)$ by using the evaluation map, $\omega \mapsto \text{ev}^*\omega$. Explicitly, we start with the $p$-form
\bea
\omega = \frac{1}{p!} \omega_{M_1 \ldots M_p}(x) dx^{M_1} \wedge \ldots \wedge dx^{M_p}
\eea
On $S^1\times LM$ the differential operator decomposes as
\bea
d &=& d_{S^1} + \delta
\eea
where
\bea
ds_{S^1} &=& ds \frac{d}{ds}\cr
\delta &=& \int ds \delta C^M(s) \frac{\delta}{\delta C^M(s)}
\eea
This means for the differentials in $\Omega^1(S^1\times LM)$ that they decompose as 
\bea
d(C^{M}(s)) &=& \frac{dC^{M}}{ds}ds + \delta C^{M}(s)
\eea
and the pullback form in $\Omega^p(S^1 \times LM)$ via the evaluation map becomes 
\bea
\text{ev}^*\omega = \frac{1}{p!} \omega_{M_1 \ldots M_p}(C(s)) \, \delta C^{M_1}(s) \wedge \ldots \wedge \delta C^{M_p}(s)\cr
+ \frac{1}{(p-1)!} \omega_{M_1 \ldots M_p}(C(s)) \, \delta C^{M_1}(s) \wedge \ldots \wedge \delta C^{M_{p-1}}(s) \wedge \frac{dC^{M_p}}{ds} ds
\eea
Here the first term is a $p$-form in $\Omega^p(LM)$ and the second term is a one-form on $S^1$ times a $(p-1)$-form on $LM$. This decomposition shows the general K\"unneth decomposition   
\bea
\Omega^p(S^1\times LM) &=& (\Omega^0(S^1) \otimes \Omega^p(LM)) \oplus (\Omega^1(S^1) \otimes \Omega^{p-1}(LM))
\eea
where $\Omega^0(M) = \C^{\infty}(M)$ is the set of smooth functions on $M$ and $\Omega^{0}(LM) = \C^{\infty}(LM)$ is the set of smooth functionals of the entire loop. 

Since $\delta C^M(s)$ and $ds$ are differentials in two different spaces, let us begin with introducing wedge products in each space separately, $\wedge_{S^1}$ that acts on $S^1$ and $\wedge_{LM}$ that acts in $LM$. We shall let the wedge product in the product manifold $S^1 \times LM$, 
\bea
\wedge &=& \wedge_{S^1} \otimes \wedge_{LM}\label{loopwedge}
\eea
act on $\alpha_1 \in \Omega^{p_1}(S^1)$, $\alpha_2 \in \Omega^{p_2}(S^1)$ and $\omega_1 \in \Omega^{r_1}(LM)$, $\omega_2 \in \Omega^{r_2}(LM)$ as
\bea
(\alpha_1 \otimes \omega_1) \wedge (\alpha_2 \otimes \omega_2) &=& (-1)^{r_1 p_2} (\alpha_1 \wedge_{S^1} \alpha_2) \otimes (\omega_1 \wedge_{LM} \omega_2)
\eea
In particular we get
\bea
(ds \otimes 1) \wedge (1 \otimes \delta C^M(s)) &=& ds \otimes \delta C^M(s)\cr
(1 \otimes \delta C^M(s)) \wedge (ds \otimes 1) &=& - ds \otimes \delta C^M(s)
\eea
We may simplify the notation and write $\wedge$ for all these products while adopting the wedge multiplication rules
\bea
\delta C^M(s) \wedge \delta C^N(s') &=& - \delta C^N(s') \wedge \delta C^M(s)\cr
ds \wedge \delta C^M(s) &=& - \delta C^M(s) \wedge ds\cr
ds \wedge ds &=& 0
\eea

A generic element in $\Omega^p(LM)$ takes the form
\bea
\omega &=& \int ds_1 \cdots \int ds_p \omega_{M_1 \cdots M_p}(s_1,\dots,s_p,C) \delta C^{M_1}(s_1) \wedge \cdots \wedge \delta C^{M_p}(s_p)
\eea
Here $\omega_{M_1 \cdots M_p}(s_1,\dots,s_p,C)$ are required to transform as a one-form with respect to each $s_k$ for $k=1,\dots,p$ to ensure reparametrization invariance of $\omega \in \Omega^p(LM)$. A vector field in loop space takes the general form 
\bea
V &=& \int ds V^M(s,C) \frac{\delta}{\delta C^M(s)}
\eea
These vector fields are reparametrization invariant if we require $V(s,C)$ to transform as a zero-form with respect to $s$. The canonical example of such a vector field is the one that is induced by a vector field $V = V^M \partial_M$ on $M$, whose loop space counterpart is 
\bea
V &=& \int ds V^M(C(s)) \frac{\delta}{\delta C^M(s)}\label{VonLM}
\eea 
If we act with $V$ on the following function in $\Omega^0(LM)$ 
\bea
f &=& \int ds f_M(C(s)) \frac{dC^M(s)}{ds}
\eea
then we get
\bea
V(f) &=& \int ds \(\partial_M f_N - \partial_N f_M\) V^M \frac{dC^N}{ds}
\eea
which is reparametrization invariant. Furthermore, 
\bea
\delta f &=& \int ds \(\partial_M f_N - \partial_N f_M\) \delta C^M(C(s)) \frac{dC^N}{ds}
\eea
From this, when combined with the formula $V(f) = \iota_V d f$ we may infer how $\iota_V$ shall act in $\Omega^p(S^1 \times LM)$ in general. Namely, it shall act by replacing $\delta C^M(s)$ with $V^M(C(s))$ at one entry in turn and summing up the contributions with alternating signs, starting from the left with a positive sign. We then have the relation
\bea
\iota_V \ev^*(\omega) &=& \ev^* (\iota_V \omega)
\eea
for a generic $\omega \in \Omega^p(M)$. 

By evaluating the $p$-form in $\Omega^p(LM)$ at $p$ different vector fields we get a real number,
\bea
\omega(V_1,\cdots,V_p) &=& \int ds_1 \cdots \int ds_p \omega_{M_1 \cdots M_p}(s_1,\dots,s_p,C)\cr
&& V_1^{M_1}(s_1,C) \cdots V_p^{M_p}(s_p,C)
\eea
To give an example of a $p$-form in $\Omega^p(S^1 \times LM) \supset \Omega^1(S^1) \otimes \Omega^{p-1}(LM)$ that does not factorize, we may start with a two-form on $M$,
\bea
\beta = \frac{1}{2} \beta_{MN}(x) dx^M \wedge dx^N \in \Omega^2(M)
\eea
Its pullback via the evaluation map is given by
\bea
\text{ev}^* \beta &=& \beta_{MN}(C(s)) \frac{dC^M}{ds} ds \wedge \delta C^N(s) \cr
&& + \frac{1}{2} \beta_{MN}(C(s)) \delta C^M(s) \wedge \delta C^N(s)
\eea
This is an element in $\Omega^2(S^1 \times LM)$ that for a generic two-form $\beta_{MN}$ does not factorize. 

It is important that independence between $\delta C^M(s)$ and $\frac{dC^M(s)}{ds} ds$ can not be obtained by imposing the transversality constraint
\bea
g_{MN}(C(s)) \frac{dC^M(s)}{ds} \delta C^N(s) &=& 0
\eea
since that would make $\delta C^M(s)$ metric-dependent and that is not something we want for a differential. (It is also an undesired property for the transgressed gauge potential and its associated Wilson line in $LM$ to depend on the spacetime metric.) Instead, independence between $\delta C^M(s)$ and $\frac{dC^M}{ds} ds$ is realized because these differentials live in different spaces, the former in $\Omega^1(LM)$ and the latter in $\Omega^1(S^1)$.

The next step towards the transgression map is to apply the interior product with the vector field 
\bea
T = \frac{\partial}{\partial s}
\eea
We then consider the following map
\bea
\omega \mapsto (-1)^{p-1} ds \wedge \iota_T \text{ev}^*\omega \in \Omega^1(S^1) \otimes \Omega^{p-1}(LM)
\eea
The final step is to integrate this over $S^1$, which gives the transgression map $\tau: \Omega^p(M) \rightarrow \Omega^{p-1}(LM)$
\bea
\tau(\omega) &=& (-1)^{p-1} \int ds \wedge \iota_T \text{ev}^*\omega 
\eea
Then one needs one more step before one can evaluate the integral, namely a prescription how to extract the integration measure from the form that is being integrated. For that we need to fix a convention for how this shall be done. We will adopt the following measure convention. For any $\Omega \in \Omega^p(LM)$ and vector fields $V_1,...,V_p$ on $LM$, we define
\bea
\int \Omega(V_1,...,V_p) &=& 0\cr
\int ds \Omega(V_1,...,V_p) &=& \int (ds \wedge \Omega)(V_1,....,V_p)\label{MeasureConv}
\eea

Explicitly, for a $p$-form $\omega = \frac{1}{p!} \omega_{M_1 \ldots M_p}(x) dx^{M_1} \wedge \ldots \wedge dx^{M_p}$, its transgression $(p-1)$-form is given by
\bea
\tau(\omega) = \frac{1}{(p-1)!} \int ds \omega_{M_1 \ldots M_p}(C(s)) \delta C^{M_2}(s) \wedge \ldots \wedge \delta C^{M_{p-1}}(s) \frac{dC^{M_p}}{ds}\label{TransgressionMap} 
\eea
The first few transgression maps are
\bea
\tau(\alpha_0) &=& 0\cr
\tau(\alpha_1) &=& \int ds \, \alpha_M(C(s)) \frac{dC^M}{ds}\cr
\tau(\alpha_2) &=& \int ds \, \alpha_{MN}(C(s)) \delta C^M(s) \frac{dC^N}{ds} 
\eea
for $p=0,1,2$ respectively. For $p=0$ the transgression map is zero since there is no fiber component $ds$ in a zero-form and therefore $\iota_T \alpha_0 = 0$.

Let us now go back to where we started. We started with 
\bea
\ev^*(\omega) &=& \omega_{M_1 \cdot M_p} \delta C^{M_1} \wedge \cdots \wedge \delta C^{M_{p-1}} \wedge \dot{C}^{M_p} ds + ...
\eea
where $+... \in \Omega^p(LM)$ that integrates to zero. The transgression map may now be defined more simply as
\bea
\tau(\omega) &=& (-1)^{p-1} \int \ev^*(\omega)\label{TraMap}
\eea
It shall be noted that the alternating sign factor here is an artifact of a convention we made for the integration measure. Other authors may use a different measure convention and accordingly have a different sign factor in (\ref{TraMap}), but at the end of the day, these transgression maps are exactly the same, and in local coordinates given by (\ref{TransgressionMap}). 

The transgression map commutes with the exterior derivative,
\bea
\tau(d\omega) = \delta(\tau(\omega))
\eea
Here $d$ on the left-hand side acts on $\Omega^p(M)$ while $\delta$ on the right-hand side acts on $\Omega^{p-1}(LM)$. 

To show this, we start with $\omega \in \Omega^p(M)$ and its pullback $\text{ev}^*\omega \in \Omega^p(S^1 \times LM)$. We introduce the operator ${\P}$ that extracts the component along the tangent direction of the loop as
\begin{equation}
{\P}(\omega) := (-1)^{p-1} ds \wedge \iota_T \text{ev}^*\omega
\end{equation}
where, as before, $T = \frac{\partial}{\partial s}$. The transgression map is obtained by integrating over the loop,
\bea
\tau(\omega) = \int \P(\omega) = (-1)^{p-1} \int_{S^1} \ev^*(\omega)
\eea
using our measure convention (\ref{MeasureConv}). We will first obtain the commutator of the exterior derivative with ${\P}$. We start with
\begin{align}
d {\P}(\omega) &= d((-1)^{p-1} ds \wedge \iota_T \ev^*\omega) \cr
&= (-1)^{p-1} ds \wedge \delta(\iota_T \ev^* \omega)
\end{align}
Now we apply Cartan's formula for the Lie derivative along $T$,
\begin{equation}
\delta (\iota_T \ev^*\omega) + \iota_T (d \:\ev^*\omega) = \mathcal{L}_T \ev^*\omega 
\end{equation}
Then we get
\begin{align}
d{\P}(\omega) &= (-1)^{p-1} ds \wedge \mathcal{L}_T \ev^*\omega - (-1)^{p-1} ds \wedge \iota_T (d\: \ev^*\omega)
\end{align}
Next we note that (by definition)
\begin{align}
{\P}(d\omega) &= (-1)^p ds \wedge \iota_T \ev^* (d\omega)
\end{align}
As a general rule the exterior derivative commutes with the pullback map. Therefore 
\bea
d{\P}(\omega) - \P (d \omega) &=& (-1)^{p-1} ds \wedge \mathcal{L}_T \ev^* \omega\label{dP}
\eea
The right-hand side is an exact term on $S^1$ -- any term involving $ds$ in the pullback $\ev^* \omega$ will be removed when we hit that with $ds \wedge$ and what remains of $\ev^* \omega$ is a form that lies entirely within $\Omega^{p-1}(LM)$ on which $\L_T$ acts as if $\ev^*\omega$ was a scalar so the right-hand side is 
\bea
(-1)^{p-1} ds \wedge \mathcal{L}_T \ev^* \omega &=& (-1)^{p-1} ds \wedge \frac{d}{ds} (\ev^* \omega)
\eea
We shall now integrate (\ref{dP}) over the loop to show that $[d,\tau] = 0$. For the second term on the left hand side of (\ref{dP}), integration immediately gives
\bea
\tau(d\omega) &=& \int \P (d\omega)
\eea
The other terms are less straightforward. Let us analyze what the first term gives,
\bea
\int d \P(\omega) = \int \delta \P(\omega) = \delta \int \P(\omega) = \delta \tau(\omega)
\eea
where in the first step we decomposed $d = d_{S^1} + \delta$ acting in respective spaces and noticed that the integration over $S^1$ will remove any $d_{S^1}(...)$ exact term, thus leaving us with a differential that acts on $LM$ which is insensitive to integration over $S^1$ and can be taken outside the integral and we arrive at $\delta$ acting on the transgressed form. We may now summarize findings. We have shown that 
\bea
d{\P}(\omega) - \P (d \omega) &=& (-1)^{p-1} ds \wedge \frac{d}{ds} \(\ev^* \omega\)
\eea
and upon integration along the loop, we get
\bea
\tau(d\omega) - \delta(\tau(\omega)) &=& 0
\eea
This completes the proof.

Let us illustrate how this works explicitly for $p = 1$. We start with the zero-form
\bea
\tau(\alpha) &=& \int ds \alpha_M(C(s)) \frac{dC^M}{ds} 
\eea
and vary the loop infinitesimally,
\bea
\delta \tau(\alpha) &=& \int ds \(\delta C^N(s) \partial_N \alpha_M \frac{dC^M}{ds} + \alpha_M \frac{d\delta C^M}{ds}\)
\eea
We make an integration by parts in the second term to get
\bea
\delta \tau(\alpha) =\int ds \(\partial_N \alpha_M - \partial_M \alpha_N\) \delta C^N \frac{dC^M}{ds} + \int ds \frac{d}{ds}\(\alpha_M \delta C^M\) = \tau (d\alpha)
\eea
Having seen that both the exterior derivative and the interior product commute with the transgression map, it follows that the Lie derivative commutes with the transgression map,
\bea
\tau(\L_V \alpha) &=& \tau((d\iota_V + \iota_V d)\alpha)\cr
&=& \(\delta \iota_V + \iota_V \delta\) \tau(\alpha)\cr
&=& \L_V \tau(\alpha)
\eea
Here we use a notation such that $V$ in $M$ acts as $V = V^M \partial_M$ and in loop space it acts according to (\ref{VonLM}). Accordingly $\L_V$, with the appropriate version of $V$, shall be chosen according to the space in which the form-field lives on which it acts. Same goes for the interior product $\iota_V$, which version of $V$ again, shall be chosen according to the space in which the form-field lives on which it acts. 

The transgression map is composed of several operations as follows,
\bea
\ev^* &:& \Omega^p(M) \rightarrow \Omega^p(S^1 \times LM)\cr
\iota_T &:& \Omega^p(S^1 \times LM) \rightarrow \Omega^{p-1}(S^1\times LM)\cr
ds \wedge &:& \Omega^{p-1}(S^1 \times LM) \rightarrow \Omega^1(S^1) \otimes \Omega^{p-1}(LM)\cr
\int_{S^1} &:& \Omega^1(S^1) \otimes \Omega^{p-1}(LM) \rightarrow \Omega^{p-1}(LM)
\eea
such that when all of these operations are combined in the order given, it gives us the transgression map $\tau: \Omega^p(M) \rightarrow \Omega^{p-1}(LM)$. Now let us begin with focusing on the first map, $\ev^*$. For a $p$-form $\alpha \in \Omega^p(M)$ we map this to $\ev^* \alpha$. The question we ask now is if we can reverse this map and get back $\alpha$? There might be various ways to get back $\alpha$ since loop space is huge. But there is one canonical way to get back $\alpha$. This is by restricting to the subspace of $LM$ consisting of constant loops. A constant loop is defined as a loop that is such that $C^M(s) = x^M$ for all $s \in S^1$. This is true regardless of how we choose the parameter $s$ and therefore does not require any gauge fixing of the parameter $s$. For a constant loop we have that the loop variation $\delta C^M(s)$ leads to the variation of the point $\delta x^M$ in $M$. Thus by promoting these to differentials on loop space and on spacetime respectively, we see that $\delta C^M(s)$ for a constant loop is nothing but the differential $dx^M$ in $\Omega^1(M)$. Furthermore, for a constant loop, $\dot{C}^M(s) := \frac{dC^M}{ds} = 0$. Thus, if we evaluate the pullback form at a constant loop, then we recover the spacetime form that we started with,
\bea
\ev^*(\alpha)(x) &=& \alpha(x)
\eea
The pullback via the evaluation map is linear, as for any real constants $a$ and $b$ we have
\bea
\text{ev}^*(a \omega_1 + b \omega_2) = a\, \text{ev}^*(\omega_1) + b\, \text{ev}^*(\omega_2)
\eea
The pullback preserves the wedge product
\bea
\text{ev}^*(\omega_1 \wedge \omega_2) = \text{ev}^*(\omega_1) \wedge \text{ev}^*(\omega_2)
\eea
and it commutes with the exterior derivative 
\bea
d\, \text{ev}^*(\omega) = \text{ev}^*(d\omega)
\eea
where on the left-hand side $d= \delta + d_{S^1}$. To show that ev$^*$ is injective we only need to note that ev$:S^1 \times LM \rightarrow M$ is surjective from which it immediately follows that ev$^*$ has no kernel.\footnote{We thank the referee for this argument.} Furthermore, ev$^*$ is surjective onto its image $\mathcal{I}^p \subset \Omega^p(S^1\times LM)$. This means that $\text{ev}^* : \Omega^p(M) \rightarrow \mathcal{I}^p$ is an isomorphism for $p = 0,1,2,...$

Since the transgression map relates $\Omega^p(M)$ to $\Omega^{p-1}(LM)$ it is natural to ask to what extent we can map operations on $\Omega^p(M)$ to corresponding operations on $\Omega^{p-1}(LM)$ via the transgression map, such as taking wedge product of forms, acting by the coadjoint differential operator on forms, taking the Hodge dual of forms, and constructing an inner product of forms. The obstruction to carry out such a programme is the fact that the transgression map is not injective. To show this, let us consider the transgression of any exact one-form on $M$,
\bea
d\omega &=& \partial_M \omega dx^M
\eea
Its transgression is a total derivative that vanishes when integrated around the loop,
\bea
\tau(d\omega) &=& \int_0^{2\pi} ds \partial_M \omega \dot{C}^M = \int_0^{2\pi} ds \frac{d\omega}{ds} = \omega(C(2\pi)) - \omega(C(0)) = 0
\eea
This is consistent with the formula $\tau(d\omega) = \delta \tau(\omega)$ since for $\omega \in \Omega^0(M)$ we have $\tau(\omega) = 0$ as a result of $\iota_V \omega = 0$. 

The only way that an integral over $S^1$ can vanish is if the integrand is exact. On $S^1$ which is a one-dimensional manifold, there are only three types of forms. The zero-forms, the harmonic one-forms and the exact one-forms. The only forms that can be integrated are the one-forms that by Hodge decomposition take the general form 
\bea
\alpha(s) ds &=& h(s) ds + d\beta(s)
\eea
as there are no co-exact one-forms in one dimension. Here $h$ is harmonic and $d\beta$ is exact. Let us assume that the metric on $S^1$ is $g_{ss}$ . Then any harmonic one-form satisfies
\bea
\(\frac{1}{\sqrt{g_{ss}}} h'(s)\)' &=& 0
\eea
whose general solution is $h(s) = h_0 + h_1 \int_0^s ds' \sqrt{g_{ss}(s')}$ but periodicity $h(s) = h(s+2\pi)$ forces $h_1 = 0$. The most general harmonic is a constant function $h(s) = h_0$. Let us now assume that
\bea
\int ds \alpha &=& 0
\eea
Then $h_0 = 0$. We find that $\alpha = d \beta$ is exact. 

Let us assume that $\omega \in \Omega^p(M)$ is closed, $d\omega = 0$. Then the transgressed form $\tau(\omega) \in \Omega^{p-1}(M)$ is also closed, $\delta \tau(\omega) = \tau(d\omega) = 0$. 

Furthermore, if $\omega$ is exact, $\omega = d\eta$ then $\tau(\omega)$ is also exact, $\tau(\omega) = \tau(d\eta) = \delta \tau(\eta)$. If we assume that $\tau(\omega)$ is closed then $0 = \delta \tau(\omega) = \tau(d\omega)$. If $\tau$ is injective, then this implies $d\omega = 0$ and so $\omega$ is closed. It is not true that $\tau$ is injective in general since when $\omega \in \Omega^0(M)$ we have $\tau(d\omega) = 0$ for any $\omega \in \Omega^0(M)$. 

We will now present an argument for that the transgression map is injective for $p = 2,3,4,...$. By linearity of the transgression map, it is sufficient to show then that if $\tau(\omega) = 0$ for $\omega \in \Omega^p(M)$ then $\omega = 0$. 

Let us pick a point $x_0$ in $M$ and $p$ linearly independent arbitrary tangent vectors $v,w_1,\dots,w_{p-1}$ in the tangent space $T_{x_0}M$ of that point. We now want to construct some convenient loops that are suitable to show injectivity. We start by introducing a smooth bump function $\phi$ on $S^1$ such that it has support in some short open interval $I_{\delta} = (-\delta,\delta)$ mod $2\pi$ for some $\delta>0$ and $\delta \less  2\pi$ where $\phi >0$ such that $\phi$ and all its derivatives vanish at the endpoints of $I_\delta$. Let us pick a point $s_0\in I_\delta$ where $\dot\phi(s_0)>0$. We then introduce a family of loops
\bea
C^M_\varepsilon(s) &=& x_0^M + \varepsilon\,\(\phi(s)-\phi(s_0)\)v^M(x_0) + \O(\epsilon^2)
\eea
where the $\O(\epsilon^2)$ corrections are there to assure the loop stays inside $M$. We can make this more precise by introducing a family of geodesics $t \mapsto \gamma^M(t,s)$ at each parameter value $s$ such that $\gamma^M(0,s) = x_0^M$ (for all $s$) and $\frac{d\gamma^M(0,s)}{dt} = \(\phi(s) - \phi(s_0)\) v^M(x_0)$. Then we define the bump loop as
\bea
C^M_{\eps}(s) &:=& \gamma^M(\eps,s)\cr
&=& x_0^M + \varepsilon\,\(\phi(s)-\phi(s_0)\)v^M(x_0) + \O(\epsilon^2)
\eea
where the first line is the exact definition that ensures that the bump loop stays in $M$ and the second line is its expansion in powers of $\eps$. This shows that the loop backtracks along the same path once it reaches the peak value of $\phi(s)$. Namely if $\phi(s_1) = \phi(s_2)$ for two distinct parameter values $s_1<s_2$ then the geodesic by the construction is such that $\gamma^M(\eps,s_1) = \gamma^M(\eps,s_2)$, which means that the loop backtracks exactly. Despite that, the loop is everywhere smooth in the sense that the embedding map $S^1 \mapsto M$ is everywhere smooth as a consequence of that the bump function is everywhere smooth. 

We may parallel transport the vector $w_j$ along the loop to obtain
$\,w_j^{\parallel}(s)\in T_{C_\varepsilon(s)}M$ such that
$w_j^{\parallel}(s_i)=w_j$ for each $s_i$ where $\phi(s_i) = \phi(s_0)$. Parallel transport along the geodesic and back along the same geodesic leaves these vectors unchanged so the transported fields return to their initial values with no holonomy jump. 

We now define vector fields in loop space along the loop as
\bea
X_1(s) &=& \dot\phi(s)\,w_1^{\parallel}(s)\cr
X_j(s) &=& w_j^{\parallel}(s)\quad (j=2,\dots,p-1)
\eea
Each vector field $X_j(s) := X_j^M(s) \frac{\delta}{\delta C^M(s)}$ is smooth along the loop. Only $X_1$ is supported in $I_\delta$ while $X_j$ for $j = 2,...,p-1$ are nontrivially supported allover the loop.

Since $\tau(\omega)=0$,
\bea
0 =\tau(\omega)(X_1,\dots,X_{p-1}) =\int ds \omega\(X_1(s),\dots,X_{p-1}(s),\dot C_\varepsilon(s)\)
\eea
We have the expansions,
\bea
C_\eps(s) &=& x_0 + \O(\eps)\cr
\dot{C}_\varepsilon(s) &=& \varepsilon\,\dot\phi(s)\,v + \O(\varepsilon^2)\cr
X_1(s) &=& \dot\phi(s)\,w_1 + \O(\varepsilon)\cr
X_j(s) &=& w_j + \O(\varepsilon)\quad(j\ge2)
\eea
which lead to
\bea
\omega\bigl(X_1,\dots,X_{p-1},\dot C_\varepsilon\bigr)
&=& \varepsilon\,\dot\phi(s)^2 \omega(x_0)(w_1,\dots,w_{p-1},v) + \O(\varepsilon^2)
\eea
Integrating over the entire loop gives the transgression that we shall put to zero by assumption,
\bea
0 &=& \varepsilon\,\omega(x_0)(w_1,\dots,w_{p-1},v)\int ds \dot\phi(s)^2 + \O(\varepsilon^2).
\eea
Here $\int\dot\phi(s)^2\,ds>0$ so we can divide by this factor. Then we may also divide by $\varepsilon$ and let $\varepsilon\to0$ to get
\bea
\omega(x_0)(w_1,\dots,w_{p-1},v) &=& 0
\eea
Since $x_0$ and $v,w_j$ are arbitrary, we get $\omega = 0$ on $M$.  This shows that $\tau$ is injective for $p\ge2$. Notice that when $p = 2$ we only have one vector field $X_1 \sim \dot{\phi}$. This also shows that this bump loop argument can not be extended to $p = 1$, at least not in a straightforward way. But on the other hand, we do not need to search for a less straightforward extension down to $p = 1$ since we have already seen that the transgression map is not injective for $p = 1$. 

As a corollary we have that the integration map over the circle when restricted to the image of the unintegrated transgression map is an isomorphism between the images of the unintegrated and the images of the integrated transgression maps for $p \geq 2$. This follows direclty from the fact that $\tau$ for $p \geq 2$ is injective. (If $\tau(\omega) = 0$ then $\omega = 0$ and consequently $\P(\omega) = 0$ so the integration map when restricted to the image of $\P$ has no kernel.)

\subsection{The unintegrated transgression map for $p \geq 0$}
From our field theory perspective it is unnatural to have this sharp distinction between $p = 1$ and $p\geq 2$. For example, we have a transgressed one-form gauge potential in $LM$ from the two-form gauge potential $B_{MN}$ on $M$ as
\bea
\tau(B) &=& \int ds B_{MN}(C(s)) \delta C^M(s) \dot{C}^N(s)
\eea
as an injective map. But we do not have an injective map for the scalar field, not even if we use a conformal Killing vector field $V^M$ and associated one-form $ V =  V_M dx^M$ as
\bea
\tau(V \phi) &=& \int ds V_M(C(s)) \phi(C(s)) \dot{C}^M(s)
\eea
One could also ask for the loop space uplift of the scalar field itself. If we apply the transgression map then this is simply vanishing,
\bea
\tau(\phi) &=& 0
\eea
These problems that arise for $p = 1$ and $p = 0$ will now be resolved by considering the unintegrated transgression map that we propose an extension for all the way down to $p = 0$, that is, to scalar fields on $M$. 

For any differential form $\omega \in \Omega^p(M)$ for $p=1,2,3,...$, we define the unintegrated transgression map $\P : \Omega^p(M) \rightarrow \Omega^1(S^1) \otimes \Omega^{p-1}(LM)$ as 
\bea
{\P}(\omega) := (-1)^{p-1} ds \wedge \iota_T \text{ev}^*\omega
\eea
Then for $p=1$ we have the map
\bea
\omega = \omega_M(x) dx^M \mapsto \P(\omega)(s,C) = ds \wedge \omega_M(C(s)) \dot{C}^M(s)
\eea
This is a one-form on the one-dimensional loop. Let us now obtain the unintegrated transgression of an exact one-form,
\bea
\partial_M \omega(x) dx^M \mapsto \P(d\omega)(s,C) = ds \wedge \frac{d}{ds} \omega(C(s)) = d_{S^1} \omega(C(s))
\eea
We thus discovered the relation
\bea
\P(d\omega) &=& d_{S^1} \ev^* \omega\label{exd}
\eea
for $\omega \in \Omega^0(M)$. In contrast to this case, for $\omega \in \Omega^p(M)$ for $p =1,2,3...$ we have
\bea
\P(d\omega) &=& \delta \P(\omega) + (-1)^p d_{S^1} \ev^* \omega\label{unified}
\eea
But in fact the relation (\ref{unified}) extends naturally to $p = 0$ since when $p = 0$ we have $\P(\omega) = 0$. 

We will introduce a weak equality $\cong$ between any two elements in $\Omega^1(S^1) \otimes \Omega^{p-1}(LM)$ for $p = 1,2,3,...$ whenever the two quantities are equal after integration over $S^1$. We will thus write a weak equality as
\bea
\alpha \cong \beta
\eea
if we have the equality
\bea
\int_{S^1} \alpha = \int_{S^1} \beta
\eea

We define a cross product of forms in the image of the transgression map im($\tau$) as the wedge product of the corresponding forms in $\Omega^*(M)$ following \cite{Hofman:2002ey},
\bea
\tau(\alpha) \times \tau(\beta) & := & \tau(\alpha \wedge \beta)
\eea
For $p = 1,2,...$, we define the inner product of two transgressed forms in $\Omega^{p-1}(LM)$ as the corresponding inner product on $M$
\bea
(\tau(\alpha),\tau(\beta))_{LM} = (\alpha,\beta)_M = \int_M \N^{p-3} \alpha \wedge *\beta\label{innerproductN}
\eea
Here $\N>0$ is an everywhere positive definite weight function on $M$. We define the inner product with the weight $\N^{p-3}$ since that makes the inner product of two Weyl invariant $p$-forms Weyl invariant. 

We define the codifferential $d^{\dag}$ as
\bea
(\alpha,d\beta)_M &=& (d^{\dag}\alpha,\beta)_M
\eea
Let us compute the left-hand side ($\alpha \in \Omega^p(M)$)
\bea
(\alpha,d\beta)_M &=& \int \N^{p-3} \alpha \wedge * d\beta\cr
&=& \int \N^{p-3} d\beta \wedge *\alpha\cr
&=& (-1)^p \int \beta \wedge d\(\N^{p-3} *\alpha\)\cr
&=& (-1)^p (-1)^{1 + (7-p)(p-1)} \int \N^{p-4} \frac{1}{\N^{p-4}} \beta \wedge ** d\(\N^{p-3} *\alpha\)\cr
&=& - \int \N^{p-4} \frac{1}{\N^{p-4}} * d \(\N^{p-3} * \alpha\) \wedge *\beta
\eea
Hence, since $\beta$ is arbitrary, we identify
\bea
d^{\dag} \alpha &=& - \frac{1}{\N^{p-4}} * d \(\N^{p-3} * \alpha\)
\eea

This definition is enabling us to move operations in loop space down to ordinary spacetime operations. For example, we have the following result for the co-differential operator $\delta^{\dag}$ in loop space,
\bea
\delta^{\dag} \tau(\alpha) &=& \tau(d^{\dag} \alpha)\label{coadj}
\eea
On the left-hand side $\delta^{\dag}$ acts on an element in $\Omega^{p-1}(LM)$ while on the right-hand side $d^{\dag}$ acts on the corresponding element in $\Omega^p(M)$. To see how to get this result, we expand the following inner product for an arbitrary $\beta \in \Omega^p(M)$,
\bea
(\delta^{\dag} \tau(\alpha),\tau(\beta))_{LM} &=& (\tau(\alpha),\delta\tau(\beta))_{LM}\cr
&=& (\tau(\alpha),\tau(d\beta))_{LM} \cr
&=& (\alpha,d\beta)_M \cr
&=& (d^{\dag}\alpha,\beta)_M \cr
&=& (\tau(d^{\dag}\alpha),\tau(\beta))_{LM}
\eea
from which (\ref{coadj}) follows. 
 
Let $V = V_M dx^M \in \Omega^1(M)$ and $\alpha \in \Omega^p(M)$ for $p = 2,3,4,...$. Then we have
\bea
\delta^{\dag} (\tau(V) \times \tau(\alpha)) &=& - \L_V \tau(\alpha) - \tau(V) \times \tau(d^{\dag} \alpha)\label{coadjapp} 
\eea
To show this relation, we first move the problem down to $M$ as
\bea
\delta^{\dag} (\tau(V) \times \tau(\alpha)) = \delta^{\dag} \tau (V\wedge \alpha) = \tau (d^{\dag} (V\wedge \alpha))
\eea
We can now carry out the computation on $M$. We consider the inner product with an arbitrary test form $\beta \in \Omega^p(M)$,
\bea
(d^{\dag} (V \wedge \alpha), \beta) &=& (V \wedge \alpha,d\beta)\cr
&=& (\alpha,\iota_V d \beta)\cr
&=& (\alpha,\L_V \beta - d \iota_V \beta)\cr
&=& - (\L_V \alpha,\beta) - (d^{\dag} \alpha,\iota_V\beta)\cr
&=& - (\L_V \alpha + V \wedge d^{\dag} \alpha,\beta)
\eea
from which follows that 
\bea
d^{\dag} (V \wedge \alpha) = - \L_V \alpha - V \wedge d^{\dag} \alpha
\eea
The final step is to transgress this result back to loop space using
\bea
\tau(\L_V \alpha) &=& \L_V \tau (\alpha)\cr
\tau(V \wedge d^{\dag} \alpha) &=& \tau(V) \times \tau(d^{\dag} \alpha)
\eea
from which the relation (\ref{coadjapp}) follows. 

Let us now repeat the computation with a one-form $\alpha \in \Omega^1(M)$. We arrive at
\bea
\delta^{\dag}(\tau(V) \times \tau(\alpha)) &=& - \L_V \tau(\alpha) - \tau(V \wedge d^{\dag}\alpha)
\eea
without any obstacles. But when we try to recast the second term as 
\bea
\tau(V \wedge d^{\dag} \alpha) &\stackrel{?}{=}
& \tau(V) \times \tau(d^{\dag} \alpha)
\eea
we meet an obstacle, namely $d^{\dag} \alpha \in \Omega^0(M)$ so its transgression vanishes, $\tau(d^{\dag}\alpha) = 0$. But the left-hand side does not necessarily vanish. Explicitly we have
\bea
V \wedge d^{\dag} \alpha = - \frac{V_M}{\N} dx^M \N^3 \nabla^N \(\frac{1}{\N^2}\alpha_N\) 
\eea
whose transgression is
\bea
\tau\(V \wedge d^{\dag} \alpha\)  &=& \int ds \frac{V_M}{\N} \(-\N^3 \nabla^N \(\frac{1}{\N^2} \alpha_N\)\) \dot{C}^M
\eea
Thus we see a breakdown of the cross product.

\subsection{The introduction of weak $(-1)$-forms}
Let us now make a new different definition of the cross product as
\bea
\P(\alpha) \times \P(\beta) &=& \P(\alpha \wedge \beta)
\eea
for any $\alpha,\beta \in \Omega^p(M)$ for $p = 1,2,...$ and examine whether we can extend this definition of the cross product to the domain where $\alpha,\beta\in \Omega^p(M)$ for $p = 0,1,2,...$. Thus our imminent challenge is to see whether we can include the case when $p = 0$. Then, in particular, for $\alpha \in \Omega^1(M)$ and hence $d^\dag \alpha \in \Omega^0(M)$, we shall clearly require that
\bea
\P(V) \times \P(d^{\dag} \alpha) &=& \P(V \wedge d^\dag \alpha)
\eea
and explicitly this reads
\bea
\frac{V_M}{\N} \dot{C}^M \times \P(d^{\dag} \alpha) &=& \frac{V_M}{\N} \(-\N^3 \nabla^N \(\frac{1}{\N^2} \alpha_N\)\) \dot{C}^M
\eea
This taken together then strongly suggests that we shall define
\bea
\P(d^{\dag} \alpha)(s,C) &:=& \ev^* (d^{\dag} \alpha)(s,C)\cr
&=& - \N^3 \nabla^N \(\frac{1}{\N^2} \alpha_N(C(s))\)
\eea
Let us then extend the range of definition of the unintegrated transgression map to include $\P: \Omega^0(M) \rightarrow  \Omega^0(S^1\times LM)$ as follows. For $\omega \in \Omega^0(M)$ we define its unintegrated transgression as
\bea
\P(\omega)(s,C) = (\ev^* \omega)(s,C) = \omega(C(s))
\eea
that we will refer to as a $(-1)$-form in loop space, or as an element in $\Omega^{-1}(LM) \subset \Omega^0(S^1\times LM)$. Note that the absence of a factor in $\Omega^1(S^1)$ makes integration of a (-1)-form over the loop unnatural. We can still integrate the (-1)-form
\bea
\int ds \omega(C(s))
\eea
of course. But the integral is not reparametrization invariant. This is to say that the $(-1)$-form is not a physical field in loop space since physical fields shall be reparametrization invariant and only depend on the shape of the loop, and not on how the loop is parametrized. The Wilson surface shall be invariant under reparametrizations, hence in particular the gauge potential in loop space is required to be reparametrization invariant on each loop that foliates such a surface. If only the gauge field is reparametrization invariant and the matter fields are not, then we will see infinitely many more matter field degrees of freedom so it becomes natural to require these fields to also have the same reparametrization invariance as the gauge field, and especially so when matter fields and the gauge field mix under supersymmetry. 

Of course, to really claim that we have a space like $\Omega^{-1}(LM)$ extending down from $\Omega^p(LM)$ for $p = 0,1,2,...$ we need to have more mathematical structures, and we do not make such a claim. For us, the introduction of the $(-1)$-form shall be used with a great deal of caution in that respect. 

However, we may confirm the following rather nice structure if we extend the definition of the differential operator $\delta$ on loop space in such a way that it acts on elements in $\Omega^{-1}(LM)$ as $\delta|_{\Omega^{-1}(LM)} := d_{S^1}$ while of course it shall act as usual as $\delta|_{\Omega^p(LM)} := \delta$ (as we defined it previously) on $\Omega^p(LM)$ for $p = 0,1,2,...$. With this extension of the loop space differential operator, we find that $\delta^2 \simeq 0$ vanishes weakly (that is $\int_{S^1} \delta^2 = 0$ vanishes strongly) when it acts on any element in $\Omega^{-1}(LM)$. The following lines of computation demonstrates this,\footnote{Let us comment that if we would attempt to define $\delta|_{\Omega^{-1}(LM)} := \delta$ then while we find $\delta^2 = 0$ strongly it would lead to a subtlety in interpreting the image space as its elements would take the form $\delta \omega(C(s)) = \partial_M \omega(C(s)) \delta C^M(s)$. These are one-forms on loop space rather than the expected zero-forms, and that is not how we like a differenial operator to act.}
\bea
\delta^2 \omega &=& \delta d_{S^1} \omega(C(s))\cr
&=& \delta \(ds \dot{C}^M(s) \partial_M \omega(C(s)\)\cr
&=& \(\partial_N \partial_M \omega - \partial_M \partial_N \omega\) ds \wedge \delta C^M(s) \dot{C}^N(s)\cr
&& - ds \wedge \frac{d}{ds} \(\partial_M \phi(C(s)) \delta C^M(s)\)\cr
&\simeq & 0
\eea
To defining $\delta|_{\Omega^{-1}(LM)} = d_{S^1}$ also ties in rather rather nicely with the relation
\bea
\P(d\omega) &=& d_{S^1} \ev^* \omega\label{exd}
\eea
for $\omega \in \Omega^0(M)$ that we may now write in the form
\bea
\P (d\omega) &=& \delta \P(\omega)
\eea
In contrast to this case, for $\omega \in \Omega^p(M)$ for $p =1,2,3...$ we have
\bea
\P(d\omega) &=& \delta \P(\omega) + (-1)^p d_{S^1} \ev^* \omega
\eea
that we may also express as a weak commutation relation
\bea
\P(d\omega) &\approx & \delta \P(\omega)
\eea
Accidentally then, for $\omega \in \Omega^0(M)$ we have $\delta \P(\omega) \approx 0$ but nevertheless, we can now express these relation in a homogenous way by saying that $\P d = \delta \P$ for all $p$-forms for $p = 0,1,2,...$. 

We also can extend the domain of the cross product to $\Omega^p(M)$ for $p = 0,1,2,...$ as follows. For $\alpha_p,\beta_p \in \Omega^p(M)$ we define
\bea
\P(\alpha_0) \times \P(\beta_0) &=& \P(\alpha_0 \beta_0)\cr
\P(\alpha_0) \times \P(\beta_p) &=& \P(\alpha_0 \beta_p), \quad (p = 1,2,...)
\eea
and we see that the cross product is a map 
\bea
\times: \Omega^{p-1}(LM) \otimes \Omega^{q-1}(LM) \rightarrow \Omega^{p+q-1}(LM)
\eea
for all $p,q = 0,1,2,...$ thus including the cases when either $p$ or $q$ or both are zero that corresponds to maps  
\bea
\times: \Omega^{-1}(LM) \otimes \Omega^{-1}(LM) \rightarrow \Omega^{-1}(LM)\cr
\times: \Omega^{-1}(LM) \otimes \Omega^0(LM) \rightarrow \Omega^{p-1}(LM)
\eea
We have introduced two maps, $\delta: \Omega^{-1}(LM) \to \Omega^0(LM)$ and $\delta^{\dag}: \Omega^0(LM) \to \Omega^{-1}(LM)$ that are given explicitly by
\bea
\delta \omega(C(s)) &=& \partial_M \omega(C(s)) \dot{C}^M(s) ds\cr
\delta^\dag \(\omega_M(C(s)) \delta C^M(s)\) &=& - \N^3 \nabla^M \(\frac{1}{\N} \alpha_M\)
\eea
All this paints a suggestive picture motivating why we like to call these objects as $(-1)$-form and denote the space of these objects as $\Omega^{-1}(LM)$. These are really just names, which may be suggestive but they could also be misleading. But in the end, they are just names for objects that we already know what they are. Namaly, the $(-1)$-forms are elements in the image of ev$^*$ when it acts on $\Omega^0(M)$ and $\Omega^{-1}(LM)$ denotes that image space. In that sense, there is nothing mysterious.

\subsection{Inner product}
We define the inner product of two arbitrary elements $\alpha,\beta\in\Omega^1(S^1) \times \Omega^{p-1}(LM)$ for $p =1,2,3,...$ as 
\bea
\(\alpha,\beta\)_{S^1 \times LM} &:=& \(\int_{S^1} \alpha,\int_{S^1} \beta\)_{LM}
\eea
That is, we define the inner product by selecting only the zero modes around the loop. We do not have an explicit definition of $(\cdot,\cdot)_{LM}$, nor do we know if such an inner product exists for every element in $LM$. We will however not use this inner product for every element in $LM$ but only for those elements that arise as a transgression of an element in $M$ as 
\bea
\int_{S^1} \alpha &=& \tau(\omega)\cr
\int_{S^1} \beta &=& \tau(\rho)
\eea
for some $\omega,\rho \in \Omega^p(M)$. Then we will define their inner product to be
\bea
\(\tau(\omega),\tau(\rho)\)_{LM} := \(\omega,\rho\)_M
\eea
This definition of the inner product can be motivated as follows. Let us consider the field strength $\P(dB)$ where $B\in\Omega^2(M)$ is the two-form gauge potential. Let us compute 
\bea
-\frac{1}{2} \(\P(B),\P(B)\)_{S^1 \times LM} &=& - \frac{1}{2}\(\tau(dB),\tau(dB)\)_{LM} = -\frac{1}{2} (dB,dB)_M
\eea
Thus with the above definitions for the inner product, we descend to the usual Maxwell action on $M$. 

The unintegrated transgression map satisfies the following weak relation for $\alpha \in \Omega^p(M)$ where $p = 1,2,3,...$ 
\bea
d^{\dag} \P(\alpha) &\cong & \P(d^{\dag} \alpha)
\eea
and when $\alpha \in \Omega^0(M)$ we have (from (\ref{exd}))
\bea
d^{\dag} \P(\alpha) &\cong & 0
\eea
For $p = 1,2,3,...$, we prove the assertion by expanding 
\bea
(d^{\dag} \P (\alpha),\P(\beta))_{S^1 \times LM} &=& (\P(\alpha),d \P(\beta))_{S^1 \times LM}\cr
&=& (\P(\alpha),[d,\P]\beta)_{S^1\times LM} + (\P(\alpha),\P(d\beta))_{S^1\times LM}\cr
&=& (\tau(\alpha),\tau(d\beta))_{LM}\cr
&=& (\alpha,d\beta)_M\cr
&=& (d^{\dag}\alpha,\beta)_M\cr
&=& (\tau(d^{\dag}\alpha),\tau(\beta))_{LM}\cr
&=& (\P(d^{\dag}\alpha),\P(\beta))_{S^1\times LM}
\eea
for arbitrary $\beta \in \Omega^{p-1}(M)$ from which the assertion follows. Here we used our definition for the inner product as an integral over the loop, that shows that the commutator term is indeed vanishing,
\bea
(\P(\alpha),[d,\P]\beta)_{S^1\times LM} &=& \(\P(\alpha),ds\wedge \frac{d}{ds} \ev^*(\beta)\)_{S^1\times LM}\cr
&:=& \(\tau(\alpha),\int ds \wedge \frac{d}{ds} \ev^*(\beta)\)_{LM} \cr
&=& 0
\eea

The exterior derivative commutes weakly with the transgression map,
\bea 
d \P &\cong& \P d
\eea
and therefore acts on a cross product as follows, (for $\alpha \in \Omega^p(M)$)
\bea
d \(\P(\alpha) \* \P(\beta)\) &=& d \P \(\alpha \wedge \beta\)\cr
&\cong& \P d \(\alpha \wedge \beta\)\cr
&=& \P \(d\alpha \wedge \beta + (-1)^p \alpha \wedge d\beta\)\cr
&\cong& d \P(\alpha) \* \P(\beta) + (-1)^p \P(\alpha) \* d \P(\beta)
\eea
In the last step we have total derivative terms cross multiplied in intermediate steps but one can convince oneself that upon summing up all terms the end result will be an equality up to an exact term despite cross multiplied exact terms in general are not exact anymore. The argument why these kind non-exact terms cancel out is as follows. If we integrate over the loop, then we consider the $\tau$ transgression map that commutes strongly with the exterior derivative. Thus if we start with the identity
\bea
\tau d \(\alpha \wedge \beta\) &=& \tau \(d\alpha \wedge \beta + (-1)^p \alpha \wedge d\beta\)
\eea
then that implies that we have the following strong relation
\bea
d \tau\(\alpha \wedge \beta\) &=& \tau \(d\alpha \wedge \beta\) + (-1)^p \tau\(\alpha \wedge d\beta\)
\eea
that we rewrite as
\bea
d \(\tau(\alpha) \* \tau(\beta)\) &=&d \tau(\alpha) \* \tau(\beta) + (-1)^p \tau(\alpha) \* d \tau(\beta)
\eea
Thus when we drop the integral, we must get the corresponding weak relation as presented above.

We define the Hodge dual as
\bea
*\P(\omega) &:=& \P(*\omega)
\eea

\subsection{Two geometric vector fields}
Let $V, U \in \Gamma(TM)$ be commuting lightlike conformal Killing vector fields on a six-dimensional Lorentzian manifold $M$ with metric $g$,
\bea
g(V, V) = 0, \quad g(U, U) = 0\cr
\L_V U = 0, \quad \L_U V = 0\cr
\L_V g = \frac{\Omega}{3} g\cr
\L_U g = \frac{\Omega^\vee}{3} g
\eea
for some smooth functions $\Omega, \Omega^\vee \in C^\infty(M)$. Under a Weyl transformation $g \rightarrow e^{2\sigma} g$ these vector fields are invariant. The flat operator maps the vector fields $V$ and $U$ to their dual one-forms $V^\flat = g(V,\cdot)$ and $U^\flat = g(U,\cdot)$ We define the rescaled Weyl invariant one-forms
\bea
\vartheta_V &:=& \frac{g(V, \cdot)}{g(V,U)}\cr
\vartheta_U &:=& \frac{g(U,\cdot)}{g(V,U)}
\eea
We define a projection of a $p$-form $\omega \in \Omega^p(M)$ onto the transverse space orthogonal to $V$ and $U$ as
\bea
\Pi(\omega) = \omega - \vartheta_V \wedge \iota_U \omega - \vartheta_U \wedge \iota_V \omega + \vartheta_V \wedge \vartheta_U \wedge \iota_V \iota_U \omega
\eea
We define 
\bea
\N &=& g(V,U)
\eea
The induced Hodge dual on the transverse space is
\bea
*_4 \omega = \frac{1}{\N} \iota_V \iota_U *_6 \omega
\eea
Their transgressions are
\bea
\V \; :=\; \P(\vartheta_V) \;=\; \frac{V_M}{\N} \dot{C}^M ds\cr
\U \;:=\; \P(\vartheta_U) \;=\; \frac{U_M}{\N} \dot{C}^M ds
\eea
We have
\bea
\L_V \vartheta_V &=& 0\cr
\L_U \vartheta_V &=& 0
\eea
whose loop space counterparts read 
\bea
\L_V \P(\vartheta_V) &=& 0\cr
\L_U \P(\vartheta_V) &=& 0
\eea
There are similar relations for $\P(\vartheta_U)$. Their exterior derivatives are
\bea
d \P(\vartheta_V) = \P(d\vartheta_V) = V_{MN} \delta C^M \dot{C}^N ds\cr
d \P(\vartheta_U) = \P(d\vartheta_U) = U_{MN} \delta C^M \dot{C}^N ds
\eea
where we define $V_{MN} = \partial_M (\vartheta_V)_N - \partial_N (\vartheta_V)_M$ and $U_{MN} = \partial_M (\vartheta_U)_N - \partial_N (\vartheta_U)_M$. 

We can form a zero-form scalar field in loop space from a (-1)-form in loop space by taking the cross product with $\P(\vartheta_V)$. A scalar field $\phi$ in the tensor multiplet transforms under Weyl transformation with Weyl weight $-2$ if we define the metric to transform with the Weyl weight $+2$. It means that the combination $\N\phi$ is Weyl invariant. From this Weyl invariant combination we form the (-1)-form 
\bea
\Phi_{-1} &:=& \P\(\N\phi\)
\eea
in loop space. This is $\N\phi$ evaluated at a point on the loop. Then we can construct two Weyl invariant zero-forms in loop space by the cross product,  
\bea
&&\Phi \;:=\; \P(\vartheta_V) \* \Phi_{-1} \;=\; V_M \phi \dot{C}^M ds\cr
&&\Phi^\vee \; :=\; \P(\vartheta_U) \* \Phi_{-1} \;=\; U_M \phi \dot{C}^M ds
\eea
For $\alpha \in \Omega^p(M)$ for $p = 1,2,3,...$ the interior product with $V$ is given by 
\bea
\iota_V \P(\alpha) &=& \P \(\iota_V \alpha\)
\eea
On the right-hand we have the usual interior product of a spacetime $p$-form that is subsequently transgressed into loop space.   
  
We will now obtain a loop space Hodge dual operation in transverse space to $U$ and $V$. At each point on $M$ these two vector fields define a two-dimensional plane $\mathcal{L} \subset TM$. We define the four-dimensional transverse space $\mathcal{T} \subset TM$ as the orthogonal complement
\bea
\mathcal{T} := \{ X \in TM \mid g(X, V) = g(X, U) = 0 \}
\eea
Let us define the transverse subspace of $p$-forms on $M$ as
\bea
\Omega^p_{\perp}(M) := \left\{ \omega \in \Omega^p(M) \;\middle|\; \iota_V \omega = 0, \; \iota_U \omega = 0 \right\}
\eea
The volume form $\text{vol}_6$ on $M$, defined by the metric $g$, decomposes into 
\bea
\text{vol}_6 = \N \, \vartheta_U \wedge \vartheta_V \wedge \text{vol}_4
\eea
where $\text{vol}_4$ is the volume form on the transverse space $ \mathcal{T}$ and the factor of $ \mathcal{N}$ ensures Weyl covariance. 

If $*_6$ denotes the Hodge star operator on $M$, then the induced Hodge star $*_4$ on the transverse space is given by
\bea
*_4 = \frac{1}{\mathcal{N}} \iota_U \iota_V *_6
\eea
To show this, we use the following definitions for the star operators. For arbitrary $\alpha,\beta\in \Omega^p(M)$, we define $*_6$ such that
\bea
\alpha \wedge *_6 \beta &=& \<\alpha,\beta\> \text{vol}_6
\eea
Here $\<\alpha,\beta\> = \frac{1}{p!} \alpha_{M_1\cdots M_p} \beta^{M_1\cdots M_p}$ in local coordinates and vol$_6$ is the volume top-form. For arbitrary $\alpha,\beta\in\Omega^p_{\perp}(M)$ we define
\bea
\alpha \wedge *_4 \beta &=& \<\alpha,\beta\> \text{vol}_4
\eea
Putting these definitions together,
\bea
\alpha \wedge *_6 \beta = \<\alpha,\beta\> \text{vol}_6 = \<\alpha,\beta\> \N \vartheta_U \wedge \vartheta_V \wedge \text{vol}_4
\eea
which leads to
\bea
\alpha \wedge \iota_U \iota_V *_6 \beta &=& \N \<\alpha,\beta\> \text{vol}_4\cr
&=& \N \alpha \wedge *_4 \beta
\eea
Since $\alpha$ and $\beta$ are arbitrary, we get
\bea
*_4 &=& \frac{1}{\N} \iota_U \iota_V *_6 
\eea
In four dimensions it is also consistent to introduce a rescaled Weyl invariant star operator as
\bea
*^W_4 &=& \N^{p-2} *_4
\eea
that maps Weyl invariant $p$-forms to Weyl invariant $(4-p)$-forms. It is easy to see that both $(*_4)^2 = (-1)^p$ and $(*_4^W)^2 = (-1)^p$. 

For $\omega \in \Omega^p_{\perp}(M)$ we define a star operator in loop space as
\bea
* \P(\omega) &:=& \P(*_4^W \omega)
\eea

\section{Bianchi identities in loop space for abelian gauge group}\label{fyra}
Let us begin with abelian gauge group. We expand the field strength
\bea
\F &=& d\A
\eea
in irreducible components as 
\bea
\F &=& h + \U \* f + \V \* g - \U \* \V \* \R
\eea
where 
\bea
\R &=& - \iota_U \iota_V \F\cr
f &=& \iota_V \F + \V \* \R\cr
g &=& \iota_U \F - \U \* \R
\eea
and $h$ are all irreducible, by which we mean that $\iota_U$ and $\iota_V$ when acting on $h$, $\R$, $f$ or $g$ vanish. The Bianchi identity
\bea
d \F &=& 0
\eea
induces several Bianchi identities on each irreducible component. To derive these identities we begin with computing the exterior derivative,
\bea
d \F &=& d h - \U \* df - \V \* dg - \U \* \V \* d\R\cr
&& + d\U \* f + d\V \* g\cr
&& - d\U \* \V \* \R + \U \* d \V \* \R 
\eea
From this result we get
\bea
\iota_V d \F &=&  \L_V h - \t{d f} + d\V \* \R\cr
&& + \V \* \(\L_V g - \L_U f - \t{d\R}\)\cr
\iota_U d \F &=& \L_U h - \t{d g} + d \U \* \R\cr
&& - \U \* \(\L_V g - \L_U f - \t{d\R}\)\cr
\iota_U \iota_V d\F &=& \L_V g - \L_U f - \t{d \R}
\eea
To be more precise, the result for $\iota_U \iota_V d\F$ is ambiguous since we can add $\V \* $(anything) as a consequence of that $U$ is lightlike. Likewise $\iota_V \iota_U d \F$ is ambiguous since we can add $\U \*$ (anything) as a consequence of that $V$ is lightlike. By demanding $\iota_U \iota_V d\F = - \iota_V \iota_U d\F$ these ambiguities are removed and we get a unique result, which is the result that we have presented above. We have the following component Bianchi identities
\bea
\t{d h} &=& 0\cr
\t{d f} &=& \L_V h + d\V \* \R\cr
\t{d g} &=& \L_U h + d\U \* \R\cr 
\t{d \R} &=& \L_U f - \L_V g
\eea

\section{Supersymmetry in loop space for abelian gauge group}\label{fem}
We define the tensor multiplet fields in loop space for abelian gauge group by the transgression map as
\bea
\A &=& \P\(\frac{1}{2} B_{MN} dx^M \wedge dx^N\)\cr
\Phi_{-1} &=& \P\(\N\phi\)
\eea
for the bosons and 
\bea
\Psi_1 &=& \P\(\frac{1}{2} \Psi_{MN} dx^M \wedge dx^N\)\cr
\Upsilon_0 &=& \P\(\Psi_M dx^M\)\cr
\chi &=& \P\(\frac{1}{2} \chi_{MN} dx^M \wedge dx^N\)\cr
\Upsilon &=& \P \(\psi_M dx^M\)\cr
\Psi_{-1} &=& \P\(\N\psi\)
\eea
for the fermions. Explicitly they are
\bea
\A &=& B_{MN} \delta C^M \dot{C}^N ds\cr
\Phi_{-1} &=& \N \phi
\eea
and 
\bea
\Psi_1 &=& \Psi_{MN} \delta C^M \dot{C}^N ds\cr
\Upsilon_0 &=& \Psi_M \dot{C}^M ds\cr
\chi &=& \chi_{MN} \delta C^M \dot{C}^N ds\cr
\Upsilon &=& \psi_M \dot{C}^M ds\cr
\Psi_{-1} &=& \N\psi
\eea
We expand
\bea
\Psi_1 &=& \chi + \V \* \Upsilon_0\cr
\Upsilon_0 &=& \Upsilon + \U \* \Psi_{-1}
\eea
Here 
\bea
\U \* \Psi_{-1} &=& U_M \psi \dot{C}^M ds
\eea
We have the supersymmetry variation 
\bea
\delta \A &=& i \Psi_1
\eea
that indices the variation 
\bea
\delta \F &=& i d\Psi_1
\eea
for the field strength. We shall now expand this result in irreducible components and begin with expanding
\bea
d\Psi_1 &=& d\chi + d\V \* \Upsilon_0 - \V \* d\Upsilon_0
\eea
Next, we compute the interior products
\bea
\iota_V d\Psi_1 &=& \L_V \chi + d\V \* \Psi_{-1} + \V \* \L_V \Upsilon + \V \* \U \* \L_V \Psi_{-1} - \V \* d\Psi_{-1}\cr
\iota_U d\Psi_1 &=& \L_U \chi - d\Upsilon - d\U \* \Psi_{-1} + \U \* d\Psi_{-1} + \V \* \L_U\Upsilon + \V \* \U \* \L_U \Psi_{-1}\cr
\iota_U \iota_V d\Psi_1 &=& \L_V \Upsilon + \U \* \L_V \Psi_{-1} - d\Psi_{-1} + \V \* \L_U \Psi_{-1}\cr
&=& \L_V \Upsilon + \L_V \Psi^\vee + \L_U \Psi - d\Psi_{-1}
\eea
Let us now obtain the supersymmetry variation of $f$. We have
\bea
\delta f &=& \iota_V \delta \F + ...\cr
&=& i \iota_V d\Psi_1 + ...\cr
&=& i \iota_V d \chi + i \iota_V \(\U \* d\V \* \Psi_{-1}\) + ...\cr
&=& i \L_V \chi + i d\V \* \Psi_{-1}
\eea
Here the dots illustrate the transverse part that we shall simply discard because $f$ is transverse. Let us next obtain the supersymmetry variation of $g$. We have
\bea
\delta g &=& \iota_U \delta \F + ...\cr
&=& i \iota_U d\Psi_1 + ...\cr
&=& i \iota_U d \chi - i d \Upsilon - i d\U \* \Psi_{-1} + ...\cr
&=& i \L_U \chi - i d\U \* \Psi_{-1} - i \t{d\Upsilon}
\eea
Let us next obtain the supersymmetry variation of $\R$. We have
\bea
\delta \R &=& - \iota_U \iota_V \delta \F\cr
&=& - i \iota_U \iota_V d\Psi_1\cr
&=& i d\Psi_{-1} - i \(\L_V \Upsilon + \L_V \Psi^\vee + \L_U \Psi\) 
\eea
In summary, and for all the fields, we have the following supersymmetry variations
\bea
\delta \Phi_{-1} &=& - i \Psi_{-1}\cr
\delta \Psi_{-1} &=& \L_V \Phi_{-1}\cr
\delta f &=& i \L_V \chi + i d\V \* \Psi_{-1}\cr
\delta g &=& i \L_U \chi - i d\U \* \Psi_{-1}\cr
&& - i \(d\Upsilon - \V \* \L_U \Upsilon - \U \* \L_V \Upsilon\)\cr
\delta h &=& i \(d\chi - \V \* \L_U \chi - \U \* \L_V \chi\) + i d\V \* \Upsilon\cr
\delta \R &=& i d\Psi_{-1} - i \(\L_V \Upsilon + \L_U \Psi + \L_V \Psi^\vee\)\cr
\delta \chi &=& - \frac{1}{2} \(f + *f\) - d\V \* \Phi_{-1}\cr
\delta \Upsilon &=& \frac{1}{2}\(\R - * h\)\cr
&& + d\Phi_{-1} - \L_U \Phi - \L_V \Phi^\vee
\eea 
We will now show that we have on-shell closure of these variations on each field. This exercise shows clearly the usefulness of introducing the $(-1)$-forms in path space. Without them we would not be able to work entirely in path space but would occasionally have to look back at our spacetime realizations of the path space fields. Utilizing the (-1)-forms enables us to carry out the closure computations entirely in path space.

\begin{itemize}
\item 
Closure on $\Phi_{-1}$,
\bea
\delta^2 \Phi_{-1} &=& - i \L_V \Phi_{-1}
\eea

\item
Closure on $\Psi_{-1}$
\bea
\delta ^2 \Psi_{-1} &=& - i \L_V \Psi_{-1}
\eea

\item
Closure on $f$,
\bea
\delta^2 f &=& - i \L_V f + \frac{i}{2} \L_V \(f-*f\)
\eea
after cancellation of the terms that involve $\Phi_{-1}$. We have on-shell closure on the selfdual equation of motion $f = *f$. 

\item
Closure on $g$ requires a lengthy computation where we use one of the Bianchi identities. First we get
\bea
\delta^2 g &=& - i d\R + \frac{i}{2} d \(\R + *h\)\cr
&& - i \L_U f + \frac{i}{2} \L_U \(f-*f\)\cr
&& + i \V \* \L_U\R - \frac{i}{2} \V \* \L_U\(\R+*h\)\cr
&& + i \U \* \L_V\R - \frac{i}{2} \U \* \L_V\(\R+*h\)\cr
&& + i \(d\L_U \Phi + d\L_V \Phi^\vee\)\cr
&& + i \V \* \L_U d\Phi_{-1} - i \V \* \L_U \L_V \Phi^\vee\cr
&& + i \U \* \L_V d\Phi_{-1} - i \U \* \L_V \L_U \Phi\cr
&& - i d\V \* \L_U \Phi_{-1} - i d\U \* \L_V \Phi_{-1}
\eea
We now use the Bianchi identity
\bea
d\R &=& \L_V g - \L_U f + \U \* \L_V \R + \V \* \L_U \R
\eea
Then we get
\bea
\delta^2 g &=& - i \L_V g\cr
&& + \frac{i}{2} d \(\R + *h\) + \frac{i}{2} \L_U \(f-*f\)\cr
&& - \frac{i}{2} \V \* \L_U\(\R+*h\)  - \frac{i}{2} \U \* \L_V\(\R+*h\)\cr
&& + i \L_U \(d\Phi + \V \* d\Phi_{-1} - d\V\*\Phi_{-1}\)\cr
&& + i \L_V \(d\Phi^\vee + \U \* d\Phi_{-1} - d\U\*\Phi_{-1}\)\cr
&& - i \L_U\L_V \(\V \* \Phi^\vee + \U \* \Phi\)
\eea
Here the three last lines are easily seen to be identically zero when we use
\bea
\Phi &=& \V \* \Phi_{-1}\cr
\Phi^\vee &=& \U \* \Phi_{-1}
\eea
We thus have on-shell closure on the selfdual equations
\bea
R &=& - *h\cr
f &=& * f
\eea

\item
Closure on $\R$ is shown as follows. First we get
\bea
\delta^2 \R &=& - i \L_V \R\cr
&& + \frac{i}{2} \L_V \(\R+*h\) + i d \L_V \Phi_{-1} - i \L_V d \Phi_{-1}
\eea
Then we have closure on the selfdual equation $R = - *h$ if we note that $\L_V d = d \L_V$. This relation is easily shown by using Cartan's formula for the Lie derivative $\L_V = d\iota_V +\iota_V d$ that holds in loop space acting on forms of any rank, including those of rank $-1$.

\item
Closure on $h$ requires the other Bianchi identity, $\L_V h - \t{d\R} + d\V \* \R = 0$, 
\bea
\delta^2 h &=& - i \L_V h \cr
&& + i \(\L_V h - \t{d\R} + d\V \* \R\)\cr
&& - \frac{i}{2} \L_V \(h+*\R\)
\eea
and thus we have closure on the selfduality equation of motion $h + * \R = 0$. 

\item
Closure on $\chi$ is easy to see. We get
\bea
\delta^2 \chi &=& - i \L_V \chi\cr
&& - \frac{i}{2} \L_V \(\chi - *\chi\) - \frac{i}{2} \(d\V - *d\V\) \* \Psi_{-1}
\eea
The terms on the second line are vanishing since both $\chi$ and $d\V$ are selfdual. That $d\V$ is selfdual is a geometrical fact and a consequence of $\eps_I$ being anti-Weyl spinors. 

\item
Closure on $\Upsilon$ is as follows,
\bea
\delta^2 \Upsilon &=& - i \L_V \Upsilon\cr
&& - \frac{1}{2} \delta \(\R + *h\)
\eea
Here we have closure on a fermionic equation of motion that is implicitly given by the supersymmetry variation of the selfdual equation of motion as
\bea
\delta \(\R + *h\) &=& 0
\eea
\end{itemize}

\section{Equations of motion in loop space for abelian gauge group}\label{sex}
In loop space the three fermionic equations of motion (\ref{firsteom}), (\ref{secondeom}) and (\ref{thirdeom}) become respectively
\bea
d^\dag \Upsilon &=& 2 \L_U \Psi_{-1}\cr
- d^\dag \chi + \L_V \Upsilon &=& d\Psi_{-1} - \L_V \Psi^\vee - \L_U \Psi + * \(d\V \* \Upsilon\)\cr
\L_U \chi &=& \t{d\Upsilon}^+
\eea
By a supersymmetry variation of the first equation we also get the scalar field equation in loop space
\bea
d^\dag \t{d\Phi_{-1}} &=& 2 \L_U \L_V \Phi_{-1}
\eea
To show this we have used $* \R = h$ together with the Bianchi identity $d h = 0$. Let us also go back to the spacetime formulation, starting with $\Phi_{-1} = \N \phi$, leading to $d\Phi_{-1} = \partial_M\(\N \phi\) \dot{C}^M ds$ and then for a zero-form, using the definition $d^\dag d\Phi_{-1} = - \N^3\nabla^M \(\frac{1}{\N^2} \partial_M\(\N\phi\)\)$. Finally extracting the longitudinal terms from $d\Phi_{-1}$ so as to descend to $\t{d\Phi_{-1}}$
\bea
\N^3 \nabla^M \(\frac{V_M}{\N^3} \L_U\(\N\phi\) + \frac{U_M}{\N^3} \L_V\(N\phi\)\) &=& 2 \L_V \L_U \(\N\phi\)
\eea
where we used $\nabla^M \(V_M/\N^3\) = 0$ and $\nabla^M \(U_M/\N^3\) = 0$ we now see that we now recover the scalar field equation in spacetime with all the right factors of $\N$.

The second equation may be obtained by a supersymmetry variation of $R = - *h$ and the third equation from a supersymmetry variation of $g = - * g$.

\section{The nonabelian transgression map}\label{sju}
Our nonabelian transgression map is closely related to the dimensional reduction map. 

We assume that there is some vector field $n^M$. We will not specify this vector field from the outset. Instead the theory may tell us if there are any restrictions we may have to impose on $n^M$. If we are given a nonabelian $p$-form in spacetime
\bea
\alpha &=& \frac{1}{p!} \alpha^a_{M_1 \cdots M_p} dx^{M_1} \wedge \cdots \wedge dx^{M_p} t_a
\eea
where $t_a$ are generators of the nonabelian gauge group, satisfying Lie algebra relations
\bea
[t_a,t_b] &=& i f_{ab}{}^c t_c
\eea
then we define the nonabelian transgression map for $p = 1,2,3,...$ as
\bea
\P(\alpha_p) \;=\; \frac{1}{(p-1)!} \int \alpha^a_{M_1 \cdots M_p} \delta C^{M_1} \wedge \cdots \wedge \delta C^{M_{p-1}} n^{M_p}(C(s)) t_a(s) ds
\eea
Here $t_a(s)$ satisfy a loop algebra
\bea
[t_a(s),t_b(s')] &=& i f_{ab}{}^c \delta(s-s') t_c(s)
\eea
The idea of using a loop algebra for the nonabelian tensor multiplet first appeared in \cite{Gustavsson:2005fp}, but here we develop this idea further. We have an integral over the loop. In the nonabelian case we have $t_a(s)$ that serve as basis elements that enable us to extract information about the integrand from the integral even when there is no differential $\delta C^M(s)$, which is the case of a zero-form in loop space. Below we will also explain why there is no (-1)-form in the nonabelian case. If we are given a Lie algebra trace such that 
\bea
\tr(t_a(s) t_b(s')) &=& \delta_{ab} \delta(s-s')
\eea
then by taking the trace of the transgressed form with $t_b(s')$, we are able to extract information about the integrand. This is different from the abelian case where we do not have the $t_a(s)$ generators. For $p=0$ this pattern breaks down since there is no index that can contract $n^M$ and we shall then define the transgression map to vanish, 
\bea
\P(\alpha_0) \;=\; 0
\eea
It means that in the nonabelian case there are no $(-1)$-forms in loop space that can exist freely outside cross products. We will define the cross product as
\bea
\P(\alpha_p) \* \P(\beta_q) &:=& \P \(\alpha_p \wedge \beta_q\)
\eea
where $\alpha_p$ is a geometic $p$-form on $M$, which means that it is abelian, whereas $\beta_q$ is a $q$-form field on $M$ taking values in the loop algebra. Notice that the cross product between two nonabelian field is rather ill-defined, as it would usually not be a product that lies in the loop algebra. We will not use the cross product two multiply two nonabelian fields but instead the wedge product $\wedge_{LM}$ as it was introduced in (\ref{loopwedge}) and which we will take over to the nonabelian setting unchanged, which is a wedge product that is defined directly on the loop space. Henceforth we will write that wedge product as $\wedge$ for brevity. We may now introduce a $(-1)$-form $\P(\beta_0)$ that can appear only inside a cross product as
\bea
\P(\alpha_p) \* \P(\beta_0) &=& \P \(\alpha_p \wedge \beta_0\)
\eea
If we use this way of writing the cross product for the scalar field, then we may introduce the $(-1)$-form $\Phi_{-1}$ in the cross product
\bea
\Phi \;=\; \V \* \Phi_{-1} \;=\; \P \(\vartheta_V \wedge \N\phi^a t_a\) \;=\; \int V_M n^M \phi^a t_a(s) ds
\eea
Our notation is such that 
\bea
t_a &=& \int t_a(s) ds
\eea
We will always make the $s$-dependence on $t_a(s)$ explicit, so there will be no confusion between $t_a(s)$ and $t_a$. 

For the gauge potential, we define
\bea
\A &=& \int B_{MN}^a \delta C^M n^N t_a(s) ds
\eea
An infinitesimal gauge parameter is defined as
\bea
\Lambda &=& \int \Lambda_M^a n^M t_a(s)ds
\eea
An infinitesimal gauge transformation is
\bea
\delta \A \;=\; D \Lambda \;=\; d\Lambda - i e [\A,\Lambda]
\eea
where this commutator is defined as
\bea
[\A,\Lambda] &=& \A \wedge \Lambda - \Lambda \wedge \A
\eea
Let us now compute the field strength,
\bea
\F &=& d\A - i e \A \wedge \A
\eea
We find that
\bea
\F &=& \frac{1}{2} \int ds H_{MNP}^a \delta C^M \wedge \delta C^N n^P t_a(s)\cr
&& + \frac{1}{2} \int \L_n B_{MN}^a \delta C^M \wedge \delta C^N t_a(s) ds
\eea
where
\bea
H_{MNP}^a &=& 3 \partial_{[M} B_{NP]}^a + \frac{3 e}{2} A_{[M}^b B_{NP]}^c f_{bc}{}^a
\eea
where the gauge potential is defined as $A_M^a = B_{MN}^a n^N$. Thus, if we put 
\bea
\L_n B_{MN}^a &=& 0
\eea
then we find that the nonabelian field strength takes the form
\bea
\F &=& \P(H)
\eea
where $H = \frac{1}{6}H^a_{MNP} t_a dx^M \wedge dx^N \wedge dx^P$.
Let us here also comment that we have $n^M A_M^a = 0$ from $A_M^a = B_{MN}^a n^N$ and therefore the gauge covariant Lie derivative equals the ordinary Lie derivative, $\h{\L_n} = \L_n - i e [n^M A_M,\bullet] = \L_n$. 
  
For the scalar field in loop space, we will define it as
\bea
\Phi &=& \int \phi^a V_M n^M t_a(s) ds
\eea
Then
\bea
D \Phi &=& \int D_M \(\phi^a V_M n^M\) \delta C^M t_a(s) ds
\eea
where we define the gauge covariant derivative as
\bea
D_M \varphi^a &=& \partial_M \varphi^a + e A_M^b \varphi^c f_{bc}{}^a 
\eea
Then we get
\bea
\iota_U D \Phi &=& \int \h{\L_U} \(\phi^a V_M n^M\) t_a(s) ds\cr
&=& \h{\L_U} \Phi
\eea
and thus we shall define 
\bea
\iota_U \Phi &=& 0
\eea
in order for the result to be consistent with the gauge covariant version of Cartan's formula,
\bea
\iota_U D + D \iota_U &=& \h{\L_U}
\eea
where $\h{\L_U}$ is the gauge covariant Lie derivative that can be obtained by just replacing all derivatives with covariant derivatives in the usual Lie derivative. 
 
Let us now establish the following expected relation in loop space, 
\bea
D \(\V \* \Phi_{-1}\) &=& d\V \* \Phi_{-1} - \V \* D\Phi_{-1}\label{minussign}
\eea
The left-hand side is computed as follows. First we compute
\bea
d\Phi &=& \int V_{MN} \N\phi^a \delta C^M n^N t_a(s) ds\cr
&& - \int \(\frac{V_M}{\N} \partial_N \(\N\phi^a\) - \frac{V_N}{\N} \partial_M \(\N\phi^a\)\) \delta C^M n^N t_a(s) ds\cr
&& + \int \L_n\(V_M\phi^a\) \delta C^M t_a(s) ds
\eea
and 
\bea
- i e [\A,\Phi] &=& e \int B_{MN}^b \phi^c V_P \delta C^M n^N n^P f_{bc}{}^a t_a(s) ds\label{commutatorterm}
\eea
This leads to the result
\bea
D \Phi &=& \int V_{MN} \N\phi^a \delta C^M n^N t_a(s) ds\cr
&& - \int \(\frac{V_M}{\N} D_N \(\N\phi^a\) - \frac{V_N}{\N} D_M \(\N\phi^a\)\) \delta C^M n^N t_a(s) ds\cr
&& + \int \h{\L_n}\(V_M\phi^a\) \delta C^M t_a(s)ds. 
\eea
For the right-hand side, we compute the two terms as follows,
\bea
\P \(\frac{1}{2} V_{MN} dx^M \wedge dx^N \N\phi^a t_a\) &=& \int V_{MN} \N\phi^a \delta C^M n^N t_a(s)
\eea
and
\bea
\P \(\V \* D\Phi_{-1}\) &=& \P\(\frac{V_M}{\N} dx^M \wedge dx^N D_N \(\N\phi^a\) t_a\)\cr
&=& \int \(\frac{V_M}{\N} D_N \(\N\phi^a\) - \frac{V_N}{\N} D_M \(\N\phi^a\)\) \delta C^M n^N t_a(s) ds
\eea
In order to match with the left-hand side, we shall take 
\bea
\h{\L_n} \(V_M \phi^a\) &=& 0
\eea
but to really show the equivalence, we need to examine the commutator terms that come from these two covariant derivatives closely. These commutator terms are
\bea
e \int \(V_M B_{NP}^b - V_N B_{MP}^b\) \phi^c f_{bc}{}^a \delta C^M n^N n^P t_a(s) ds
\eea
and now we can see that the first term vanishes because $B_{NP}^a$ is antisymmetric and contracted with $n^N n^P$ that is symmetric. So the commutator term is in precise agreement, including the sign once we notice the minus sign in (\ref{minussign}), with (\ref{commutatorterm}).

Let us now compute $D^\dag\(\V \* \alpha\)$ using the definition of $D^\dag$ from a putative inner product using some simple rules that this inner product is supposed to satisfy, such as
\bea
(\beta,D\alpha) &=& (D^\dag\beta,\alpha)\cr
(\beta,\V\*\alpha) &=&(\iota_V\beta,\alpha)
\eea
Then we get
\bea
\(\beta,D^\dag \(\V \* \alpha\)\) &=& (D\beta,\V \* \alpha)\cr
&=& \(\iota_V D\beta,\alpha\)\cr
&=& \(\h{\L_V}\beta,\alpha\) - \(D\iota_V \beta,\alpha\)\cr
&=& - \(\beta,\h{\L_V}\alpha\) - \(\iota_V \beta,D^\dag\alpha\)\cr
&=& - \(\beta,\h{\L_V}\alpha + \V \* D^\dag \alpha\)
\eea
where 
\bea
\h{\L_V} &=& D\iota_V + \iota_V D
\eea 
is the covariant Lie derivative that is computed as follows,
\bea
\h{\L_V} \alpha &=& \L_V \alpha - i e [\A,\iota_V \alpha] - ie \iota_V [\A,\alpha]\cr
&=& \L_V \alpha - i e [\iota_V \A,\alpha]
\eea
From the above result we read off
\bea
D^\dag \(\V \* \alpha\) &=& - \h{\L_V} \alpha - \V \* D^\dag \alpha
\eea
It is now interesting to examine what we get when 
\bea
\alpha &=& \int \alpha^a_M n^M t_a(s) ds
\eea
is a zero-form. Then 
\bea
\V \* \alpha &=& \P \(\frac{V_M}{\N} dx^M \wedge \alpha^a_N dx^N t_a\)\cr
&=& \int \(\frac{V_M}{\N} \alpha^a_N - \frac{V_N}{\N} \alpha^a_M\) \delta C^M n^N t_a(s) ds
\eea
For a $p$-form
\bea
\alpha_p &=& \frac{1}{p!} \int \alpha_{M_1 \cdots M_p M_{p+1}} \delta C^{M_1} \wedge\cdots \wedge \delta C^{M_p} n^{M_{p+1}} t_a(s) ds
\eea
we shall define the codifferential operator as
\bea
d^\dag \alpha_p &=& - \frac{1}{(p-1)!} \int \N^{3-p} D^{M_1} \(\frac{1}{\N^{2-p}} \alpha_{M_1 \cdots M_p M_{p+1}}\) \cr
&& \delta C^{M_2} \wedge \cdots \wedge \delta C^{M_p} n^{M_{p+1}} t_a(s) ds
\eea
when $p = 1,2,3,...$ and 
\bea
d^\dag \alpha_0 &=& - \N^3 D^M \(\frac{1}{\N^2} \alpha^a_M\) t_a
\eea
when $p = 0$. To see this, we repeat essentially the computation that we did in the abelian case,
\bea
D^\dag \(\V \* \alpha\) &=& - \int \N^2 D^M \(\(\frac{V_M}{\N^2} \alpha^a_N - \frac{V_N}{\N^2} \alpha^a_M\)\) t_a(s) n^N ds\cr
&=& - \int \h{\L_V} \alpha^a_N n^N t_a(s) ds + \int V_N \N^2 D^M \(\frac{1}{\N^2} \alpha^a_M\) n^N t_a(s) ds
\eea

Let us now check that having $\Phi_{-1} = 0$ is consistent with having $D\Phi_{-1} = 0$ as well. Direct application of the definition yields
\bea
D\Phi_{-1} \;=\; \P\(D_M\(\N\phi^a\) dx^M t_a\) \;=\; \int D_M\(\N\phi^a\) n^M t_a(s) ds
\eea
which vanishes only if we impose the constraint
\bea
\h{\L_n}\(\N\phi^a\) &=& 0
\eea
Hence, in the nonabelian case we have to perform dimensional reduction along $n^M$ in order for our nonabelian transgression map to be consistent. 

Let us notice that it is sufficient for us to assume that $n^M$ is a conformal Killing vector, not necessarily a Kiling vector. That is so because we require $\h{\L_n} = 0$ when it acts on loop space fields and all our loop space fields are Weyl invariant. Alternatively, if we translate back to spacetime fields then the field quantity that $\h{\L_n}$ will act on and set  to zero by the dimensional reduction, is always necessarily a Weyl invariant combination such as $\N\phi^a$.

In loop space with nonabelian gauge group we have $p$-form fields $\alpha_p$ that can be either bosonic or fermionic and they can be either wedge multiplied or cross multiplied but for the sake of illustration let us assume that they are bosonic. In commutators we always use the wedge product and never the cross product,
\bea
[\alpha_p,\beta_q] &=& \alpha_p \wedge \beta_q - \beta_q \wedge \alpha_p \qquad {\mbox{if $(-1)^{pq} = 1$}}\cr
\{\alpha_p,\beta_q\} &=& \alpha_p \wedge \beta_q + \beta_q \wedge \alpha_p \qquad {\mbox{if $(-1)^{pq} = -1$}}
\eea
Let us introduce a graded commutator
\bea
[\alpha_p,\beta_q\} &=& \alpha_p \wedge \beta_q - (-1)^{pq} \beta_q \wedge \alpha_p
\eea
with the property
\bea
[\alpha_p,\beta_q\} &=& (-1)^{pq} [\beta_q,\alpha_p\}
\eea
We have also introduced a cross product. In all our equations the cross product only appears between a geometrical field and a field in the tensor multiplet. The geometrical field does not take values in the loop algebra of the gauge group. Let us denote the geometrical field as $\alpha_r$ of rank $r$, it could be for instance $\V$ or $d\V$. Then the cross product has the following property,
\bea
[\alpha_r \* \beta_p,\gamma_q\} &=& \alpha_r \* [\beta_p,\gamma_q\}\label{axiom}
\eea
From these properties we can derive the additional property
\bea
[\alpha_r \* \beta_p,\gamma_q\} &=& (-1)^{(r+1)p} [\beta_p,\alpha_r \* \gamma_q\}
\eea
This property can be shown as follows,
\bea
[\alpha_r \* \beta_p,\gamma_q\} &=& \alpha_r \* [\beta_p,\gamma_q\}\cr
&=& (-1)^{pq} \alpha_r \* [\gamma_q,\beta_p\}\cr
&=& (-1)^{pq} [\alpha_r \* \gamma_q,\beta_p]\cr
&=& (-1)^{(r+1)p} [\beta_p,\alpha_r \* \gamma_q]
\eea

\section{Bianchi identities in loop space with nonabelian gauge group}\label{atta}
We now turn to the case of a nonabelian gauge group. The gauge field strength 
\bea
\F &=& d\A - i e \A \wedge \A
\eea
satisfies the Bianchi identity 
\bea
D \F &=& 0
\eea
that we may expand as
\bea
d \F &=& i e \(\A \wedge \F - \F \wedge \A\)
\eea
All that goes into showing this Bianchi identity are the properties 
\bea
d \(\alpha_p \wedge \beta_q\) &=& d \alpha_p \wedge \beta_q + (-1)^p \alpha_p \wedge d\beta_q\cr
(\alpha \wedge \beta) \wedge \gamma &=& \alpha \wedge \(\beta\wedge \gamma\)
\eea
that we shall require of the wedge product in loop space. Then the proof of the nonabelian Bianchi identity is just one line,
\bea
d\F \;=\; - i e \(d\A \wedge \A - \A \wedge d\A\) \;=\; - i e \(\F \wedge \A - \A\wedge \F\)
\eea
As in the abelian case, we will again expand 
\bea
\F &=& h + \U \* f + \V \* g - \U \* \V \* \R
\eea
in irreducible components such that 
\bea
\iota_V \F &=& f - \V \* \R\cr
\iota_U \F &=& g + \U \* \R\cr
\iota_U \iota_V \F &=& - \R
\eea
We then again find that
\bea
\iota_V d \F &=&  \L_V h - \t{d f} + d\V \* \R\cr
&& + \V \* \(\L_V g - \L_U f - \t{d\R}\)\cr
\iota_U d \F &=& \L_U h - \t{d g} + d\U \* \R\cr
&& - \U \* \(\L_V g - \L_U f - \t{d\R}\)\cr
\iota_U \iota_V d\F &=& \L_V g - \L_U f - \t{d \R}
\eea
What is new here are the commutator terms. First we expand $\A = \t\A + \V \* \iota_U \A + \U \* \iota_V \A$ and then we expand
\bea
\iota_V \(\A \wedge \F\) &=& \iota_V \A \wedge \F - \A \wedge \iota_V \F\cr
&=& \iota_V \A \wedge h - \t\A \wedge f\cr
&& + \iota_V \A \wedge \V \* g - \V \* \iota_U \A \wedge f + \t\A \wedge \V \* \R\cr
&& + \iota_V \A \wedge \U \* f - \U \* \iota_V \A \wedge f\cr
&& - \iota_V \A \wedge \U \* \V \* \R + \U \* \iota_V \A \wedge \V \* \R
\eea
We now apply the rule 
\bea
[\U \* \alpha_p,\beta_q] &=& (-1)^p [\alpha_p,\U \* \beta_q]\label{p}
\eea
where the commutator acts only on Lie algebra generators, alternatively is is sometimes an anticommutator depending on $p$ and $q$ (when $pq$ is odd) such that we can use the loop algebra in the end for the generators. Either way, we have 
\bea
[\U \* \alpha_p,\beta_q] &=& \int \frac{U_M}{\N} \alpha^a_{M_1\cdots M_p P} \beta^b_{N_1 \cdots N_q Q} n^P n^Q i f_{ab}{}^c t_c(s)\cr
&& \delta C^M \delta C^{M_1} \cdots \delta C^{M_p} \delta C^{N_1} \cdots \delta C^{N_q} ds\cr
[\alpha_p,\U \* \beta_q] &=& \int \alpha^a_{M_1 \cdots M_p P} \frac{U_M}{\N} \beta^b_{N_1\cdots N_q Q} n^P n^Q i f_{ab}{}^c t_c(s) \cr
&& \delta C^{M_1} \cdots \delta C^{M_p} \delta C^M \delta C^{N_1} \cdots \delta C^{N_q}  ds
\eea
Now by permuting $\delta C^M$ to the left through $p$ differentials in the second expression, we get the sign factor $(-1)^p$. We also have 
\bea
\U \* [\alpha_p,\beta_q] &=& [\U \* \alpha_p,\beta_q]
\eea
Then 
\bea
\iota_V [\A,\F] &=& [\iota_V \A,h] - [\t\A,f]\cr
&& + \V \* \([\iota_V \A,g] - [\t\A,R] - [\iota_U,f]\)
\eea
Similarly we have
\bea
\iota_U [\A,\F] &=& [\iota_U,h] - [\t\A,g]\cr
&& + \U \* \([\iota_U,f] + [\t\A,\R] - [\iota_V \A,g]\)
\eea
We can now see that the terms combine into 
\bea
\iota_V D\F &=& \h{\L_V h} - \t{D f} + d\V \* \R\cr
&& + \V \* \(\h{\L_V g} - \h{\L_U f} - \t{D \R}\)
\eea
and
\bea
\iota_U D\F &=& \h{\L_U} h - \t{D g} + d\U \* \R\cr
&& + \U \* \(\h{\L_U} f - \h{\L_V} g + \t{D\R}\)
\eea
Both these results lead to
\bea
\iota_U \iota_V D\F &=& \h{\L_V} g - \h{\L_U} f - \t{D\R}
\eea
From these relations, we conclude that we have the following nonabelian Bianchi identities,
\bea
\h{\L_V h} - \t{D f} + d\V \* \R &=& 0\cr
\h{\L_U h} + \t{D g} + d\U \* \R &=& 0\cr
\h{\L_V g} - \h{\L_U} f - \t{D\R} &=& 0
\eea

\section{Supersymmetry in loop space with nonabelian gauge group}\label{nio}
We will now construct the nonabelian generalization of the supersymmetry variations in loop space and show that they close on-shell. For the nonabelian generalization, we shall replace the differential $d$ acting on a $p$-form $\alpha$ with the covariant differential operator
\bea
D \alpha &=& d \alpha - i e [\A,\alpha]
\eea
and we will accordingly replace $\L_U \alpha$ and $\L_V \alpha$ with 
\bea
\J_V \alpha &=& \L_V \alpha - i e [\B,\alpha]\cr
\J_U \alpha &=& \L_U \alpha - i e [\C,\alpha]
\eea
where
\bea
\B &=& \iota_V \A - \Phi\cr
\C &=& \iota_U \A
\eea
Here we prefer to use $\B$ in place of $\iota_V \A$ because $\B$ will be assumed to be a supersymmetry invariant also in the nonabelian generalization. This will simplify our computations of the closure as $\bar{\L}_V$ will be invariant. However $\bar{\L}_U = \h{\L}_U$ will vary under supersymmetry. We will assume that 
\bea
\delta \C &=& i \Upsilon_0
\eea
which induces a corresponding variation of $\bar{\L}_U$. These Lie derivatives satisfy the commutation relation
\bea
[[\J_U,\J_V],\alpha] &=& - i e [\L_U \B - \L_V \C - i e [\C,\B],\alpha]
\eea
We also have
\bea
[D,\J_U] \alpha &=& i e [\iota_U \F,\alpha]\cr
[D,\J_V] \alpha &=& i e [\iota_V \F,\alpha] + i e [D\Phi,\alpha]
\eea 
Let us also note that 
\bea
[\h{\L_U},\h{\L_V}] \alpha &=& - i e [\R,\alpha]
\eea
When generalizing the abelian supersymmetry variations, we replace $d$ with $D$ and we replace $\L_V$ and $\L_U$ with $\J_V$ and $\J_U$. We then adjust additional commutators between scalar and fermion fields so that the supersymmetry variations close on-shell. In this way, we have found the following supersymmetry variations,
\bea
\delta (\alpha \* \Phi_{-1}) &=& - i \alpha \* \Psi_{-1}\cr
\delta (\alpha \* \Psi_{-1}) &=& \alpha \* \J_V \Phi_{-1}\cr
\delta \B &=& 0\cr
\delta \C &=& i \Upsilon_0\cr
\delta \t{\A} &=& i \chi\cr
\delta f &=& i \J_V \chi + i d\V \* \Psi_{-1}\cr
\delta g &=& i \J_U \chi - i d\U \* \Psi_{-1} - i \t{D\Upsilon} + e [\Phi^\vee,\chi]\cr
\delta h &=& i \t{D\chi} + i d\V \* \Upsilon\cr
\delta \R &=& - i \(\J_V \Upsilon + \J_U \Psi + \J_V \Psi^\vee\) + e [\Upsilon_0,\Phi]\cr
\delta \chi &=& - \frac{1}{2} \(f + *f\) - d\V \* \Phi_{-1}\cr
\delta \Upsilon &=& \frac{1}{2}\(\R - * h\) - \J_U \Phi - \J_V \Phi^\vee
\eea
Here $\alpha$ could be any supersymmetry invariant $p$-form satisfying $\J_V \alpha = 0$, it could for example be $\V$ or $\U$. It could also be $d\V$ or $d\U$. We have to introduce such an $\alpha$ because $\Phi_{-1} = 0$ unless it is cross multiplied with something. From the above variations one can show that the gauge potential $\A = \t\A + \V \* \iota_V \A + \U \* \iota_U \A$ will have the supersymmetry variation 
\bea
\delta \A &=& i \Psi_1
\eea
which implies   
\bea
\delta \F &=& i D\Psi_1
\eea
The supersymmetry variations of the components $f$ and $g$ involve the scalar field through commutators, but there is no scalar field in the supersymmetry variation of $\F$. There is however no contradiction here as one can add to the supersymmetry variation of $\R$ the following vanishing term
\bea
\delta \R &=& ... + e [\Phi_{-1},\chi]
\eea
This term vanishes by $\Phi_{-1} = 0$ but when cross multiplied with either $\V$ or $\U$ it precisely cancels the dependence on the scalar field, so that
\bea
\delta \iota_V \F &=& \delta f - \V \* \delta \R\cr
\delta \iota_U \F &=& \delta g + \U \* \delta \R
\eea
do not involve the scalar field. Alternatively, as one reconstructs $\F$ from the components, by adding the vanishing term to $\delta \R$ one again finds that it cancels all dependence of the scalar field in the variation $\delta \F$, hence avoiding a contradiction. 

The supersymmetry variation of $\R$ can also be expressed as
\bea
\delta \R &=& i \(\t{D\Psi_{-1}} - \h{\L}_V \Upsilon\)
\eea
showing explicitly that it does not involve the scalar field.

We have already seen that in the nonabelian case $\Phi$ is transverse. Explicitly
\bea
\Phi &=& \int \phi^a V_M n^M t_a(s) ds\cr
\Phi^\vee &=& \int \phi^a U_M n^M t_a(s)
\eea
From $\Phi$ we can extract the integrand
\bea
\phi^a V_M n^M
\eea
and since $V_M$ and $n^M$ are geometrical vector fields, we will assume that they are known to us. So the multiplicative factor $V_M n^M$ is known. That means that from $\Phi$ we can extract the value $\phi^a$ at each point along a given loop. Similarly, from $\Phi^\vee$ again we can extract $\phi^a$ at each point. 

Let us now turn to the closure computation.

\begin{itemize}
\item
Closure on $\alpha \* \Phi_{-1}$, 
\bea
\delta^2 (\alpha \* \Phi_{-1}) &=& - i \alpha \* \J_V \Phi_{-1}\cr
&=& - i \J_V \(\alpha \* \Phi_{-1}\) + i \(\J_V \alpha\) \* \Phi_{-1}
\eea
We have closure for any $\alpha$ such that $\J_V \alpha = 0$.
 
\item
Closure on $\alpha \* \Psi_{-1}$ is simplified by that we use the supersymmetric invariant $\B$ in place of $\iota_U \A$ when we define $\J_V$,
\bea
\delta^2 (\alpha \* \Psi_{-1}) &=& - i \alpha \* \J_V \Psi_{-1}\cr
&=& - i \J_V \(\alpha \* \Psi_{-1}\)
\eea
for $\J_V \alpha = 0$. 

\item
Closure on $f$ is as follows,
\bea
\delta^2 f &=& - i \J_V f + \frac{i}{2} \J_V \(f - *f\) - i \(\L_V d\V\) \* \Phi_{-1}
\eea
We can see that the last term vanishes by $\L_V d\V = d\L_V \V = 0$ and  we have closure on the selfduality equation of motion 
\bea
f &=& *f
\eea

\item
Closure on $\chi$,
\bea
\delta^2 \chi &=& - i \J_V \chi - i d\V \* \Psi_{-1} - d\V \* \(-i \Psi_{-1}\)\cr
&=& - i \J_V \chi
\eea
This could also appear to settle the issue of weather the variation of $\chi$ shall involve $d\V \* \Phi_{-1}$ or $\t{D\Phi}$. However, these expressions turn out to be identical since in the expansion
\bea
D\Phi &=& d\V \* \Phi_{-1} - \V \* D\Phi_{-1} 
\eea
when we project to the transverse space, the second term shall be dropped, so we have the identity
\bea
\t{D\Phi} &=& d\V \* \Phi_{-1}
\eea
Let us show this in more detail using our explicit realization
\bea
- \V \* D\Phi_{-1} &=& - \int \(\frac{V_M}{\N} D_N(\N\phi^a) - \frac{V_N}{\N} D_M(\N\phi^a)\) \delta C^M n^N t_a(s) ds
\eea
This implies 
\bea
\iota_V \(- \V \* D\Phi_{-1}\) &=& \int \frac{V_N n^N}{\N} \L_V \(\N\phi^a\) t_a(s) ds\cr
\iota_U \(- \V \* D\Phi_{-1}\) &=& \int \frac{V_N n^N}{\N} \L_U \(\N\phi^a\) t_a(s) ds \cr
&& - \int n^N D_N\(\N\phi^a\) t_a(s) ds
\eea
Now we use
\bea
\L_V \(\frac{V_N}{\N}\) &=& 0\cr
\L_U \(\frac{V_N}{\N}\) &=& 0
\eea
together with 
\bea
\L_V n^N \;=\; - \L_n V^N \;=\; 0\cr
\L_U n^N \;=\; - \L_n U^N \;=\; 0
\eea
and we assume that
\bea
\h{\L_n} \(\N\phi^a\) &=& 0
\eea
Then we get
\bea
\iota_V D\Phi &=& \h{\L_V} \Phi\cr
\iota_U D\Phi &=& \h{\L_U} \Phi
\eea
This shows that $\iota_V \Phi = 0$ and $\iota_U \Phi = 0$ in order to satisfy Cartan's formula for the Lie derivative. Next, we have
\bea
\iota_V \(d\V \* \Phi_{-1}\) &=& 0\cr
\iota_U \(d\V \* \Phi_{-1}\) &=& 0
\eea
From these results we see that $- \V \* D\Phi_{-1}$ will be projected out upon a projection to the transverse directions.

\item
Closure on $g$ requires the use of the Bianchi identity
\bea
\t{D \R} &=& \J_V g - \J_U f - i e [\Phi,g]
\eea
We start with 
\bea
\delta g &=& i \J_U \chi - i \t{D \Upsilon} - i d\U \* \Psi_{-1} + e [\Phi^\vee,\chi]\cr
\delta \Upsilon &=& \R - \J_U \Phi - \J_V \Phi^\vee\cr
\delta \chi &=& - f - d\V \* \Phi_{-1}
\eea
and make a second variation
\bea
\delta^2 g &=& i \J_U \delta \chi - i \t{D\delta \Upsilon} - i d\U \* \delta \Psi_{-1}\cr
&& + e \U \* \{\delta \Phi_{-1},\chi\} + e \U \* [\Phi_{-1},\delta\chi]\cr
&& + i \(\delta \J_U\) \chi - i \t{\(\delta D\) \Upsilon}
\eea
Let us begin by analyzing the fermionic terms that arise from the third line,
\bea
 i \(\delta \J_U\) \chi - i \t{\(\delta D\) \Upsilon} &=& e \{\delta \C,\chi\} - e \{\delta \t\A,\Upsilon\}\cr
&=& i e \{\Upsilon_0,\chi\} - i e \{\chi,\Upsilon\}\cr
&=& i e \{\Psi^\vee,\chi\}
\eea
This term is subsequently canceled by the term
\bea
e \U \* \{\delta \Phi_{-1},\chi\} &=& - i e \U \* \{\Psi_{-1},\chi\}
\eea
We are left with only the first line, and one commutator term,
\bea
\delta^2 g &=& - i \J_U f - i d\V \* \J_U \Phi_{-1} - i d\U \* \J_V \Phi_{-1}\cr
&& - i \t{D\R} + i \t{D \J_U \Phi} + i \t{D \J_V \Phi^\vee}\cr
&& + e \U \* [\Phi_{-1},\delta\chi]
\eea
We apply the Bianchi identity to $D\R$,
\bea
\delta^2 g &=& - i \J_V g - e [\Phi,g]\cr
&& + i \t{[D,\J_U] \Phi} + i \t{[D,\J_V] \Phi^\vee}\cr
&& + i \t{\J_U \(D\Phi - d\V \* \Phi_{-1}\)}\cr
&& + i \t{\J_V \(D\Phi^\vee - d\U \* \Phi_{-1}\)}\cr
&& + e \U \* [\Phi_{-1},\delta\chi]
\eea
The commutators from the second line are
\bea
-e[f, \Phi^\vee]-e [g, \Phi]-e[\t{D \Phi},\Phi^\vee] &=& -e [g, \Phi]+e[\delta \chi,\Phi^\vee].
\eea
We now see that the commutator terms $[\delta \chi,\Phi^\vee]$ and $[\Phi,g]$ cancel precisely, and if we also use 
\bea
\t{D \Phi} &=& d\V \* \Phi_{-1}\cr
\t{D \Phi^\vee} &=& d\U \* \Phi_{-1}
\eea
then we are left with the simple closure relation
\bea
\delta^2 g &=& - i \J_V g 
\eea
All that went into it was only just one Bianchi identity that holds off-shell.

\item
Closure on $\R$ requires the use of the commutation relation
\bea
[\J_U,\J_V]\alpha &=& - i e [\L_U \B - \L_V \C - i e [\C,\B],\alpha]
\eea
To show closure on $\R$ we start from the variation
\bea
\delta \R &=& i D \Psi_{-1} - i \(\J_U \Psi + \J_V \Psi^\vee + \J_U \Upsilon\)\cr
&& + e [\Upsilon_0,\Phi]
\eea
and make a second variation using that 
\bea
\delta \Psi &=& \J_V \Phi\cr
\delta \Psi^\vee &=& \J_V \Phi^\vee\cr
\delta \Upsilon_0 &=& \L_U \B - \L_V \C - i e [\C,\B]\cr
&=& \R - \J_U \Phi\cr
\delta \C &=& i \Upsilon_0
\eea
Then we get
\bea
\delta^2 \R &=& - i \J_V \R 
- i [\J_U,\J_V] \Phi + i [D,\J_V] \Phi_{-1}\cr
&& + e [\L_U \B - \L_V \C - i e [\C,\B],\Phi] \cr
&& - e [\delta \C,\Psi] - e [\Upsilon_0,\delta\Phi]
\eea
where we remembered that $\delta$ anticommutes with $\Upsilon_0$. Now by evaluating this, one finds cancelation among all terms except the first, so one has the closure relation
\bea
\delta^2 \R &=& - i \J_V \R - e [f,\Phi_{-1}]
\eea
The commutator term vanishes by $\Phi_{-1} = 0$ in the nonabelian case.  Hence we have closure,
\bea
\delta^2 \R &=& - i \J_V \R
\eea

\item
Closure on $\Upsilon$ is straightforward using $\delta \C = \Upsilon_0$ and then we almost immediately arrive at
\bea
\delta^2 \Upsilon &=& - i \J_V \Upsilon
\eea
\end{itemize}
This completes the check of closure of these supersymmetry variations. 

Let us now comment that we have not listed the supersymmetry variation of $h$. This is not necessary since $h$ is related by selfduality to $\R$ so its on-shell closure follows directly from the on-shell closure on $\R$. However, we may still discuss the supersymmetry variation of $h$ separately. Let us for now assume that there is an equation of motion 
\bea
* \t{D\chi} &=& \h{\L}_V \Upsilon - \t{D\Psi_{-1}} - * \(d\V \* \Upsilon\) + \xi\label{xi}
\eea
where $\xi$ can be some commutator terms. We also have the supersymmetry variation
\bea
\delta *\R &=& i * \(\t{D\Psi_{-1}} - \h{\L_V} \Upsilon\)
\eea
If we now use the equation of motion above together with $h = - * \R$, then we conclude that $h$ shall have the supersymmetry variation
\bea
\delta h &=& i \t{D\chi} + i d\V \* \Upsilon - i * \xi
\eea
Closure requires
\bea
\delta \xi &=& i e *[\Phi,h]
\eea
which on-shell is 
\bea
\delta \xi &=& - i e [\Phi,\R]
\eea
Let us next make a supersymmetry variation of the equation of motion. For the left-hand side we have the supersymmetry variation
\bea
\delta (l.h.s) &=& - * \t{D f} - * \(d\V \* \t{D \Phi_{-1}}\)
\eea
and for the right-hand side we have
\bea
\delta (r.h.s) &=& \h{\L}_V \(\R - \h{\L}_U \Phi - \h{\L}_V \Phi^{\vee}\)\cr
&& - e \{\Psi,\Upsilon\}\cr
&& - \t{D \h{\L_V} \Phi_{-1}}\cr
&& - * \(d\V \* \R\)\cr
&& + * \(d\V \* \h{\L_U} \Phi\) + * \(d\V \* \h{\L_V}\Phi^{\vee}\)\cr
&& - e \{\Upsilon,\Psi\}\cr
&& + \delta \xi
\eea
By noting the Bianchi identity of the disguised form
\bea
- * \t{D f} &=& \h{\L_V} \R - * \(d\V \* \R\)
\eea
we find that the supersymmetry transformation leads to the equation
\bea
0 &=& - \t{[D,\h{\L_V}] \Phi_{-1}} + \delta \xi
\eea
Let us expand the commutator carefully,
\bea
\t{[D,\h{\L_V}] \Phi_{-1}} &=& i e [\iota_V \F,\Phi_{-1}]\cr
&=& i e [f - \V \* \R,\Phi_{-1}]\cr
&=& i e \V \* [\R,\Phi_{-1}]\cr
&=& - i e \V \* [\Phi_{-1},\R]\cr
&=& - i e [\Phi,\R]
\eea
Hence we found that the equation of motion is supersymmetric provided  
\bea
\delta \xi &=& - i e [\Phi,\R]
\eea
We thus find the same condition for the supersymmetry variation of $\xi$ as we found above. 

Let us now show that the equation of motion
\bea
\h{\L_U} \chi &=& \t{D\Upsilon}^+ + i e [\Phi^\vee,\chi]
\eea
is supersymmetric. We have
\bea
\delta (l.h.s) &=& \h{\L_U} \delta \chi - i e \{\delta \C,\chi\}\cr
&=& - \h{\L_U} f - d\V \* \h{\L_U} \Phi_{-1} + e \{\Upsilon,\chi\} + e\{\Psi^\vee,\chi\}
\eea
and 
\bea
\delta (r.h.s) &=& \t{D\delta\Upsilon}^+ - i e [\delta\t{\A},\Upsilon]^+ \cr
&& + i e [\delta \Phi^\vee,\chi] + i e [\Phi^\vee,\delta \chi]\cr
&=& \t{D\R}^+ - \t{D\h{\L_U}\Phi}^+ - \t{D\h{\L_V}\Phi^\vee}^+ + e \{\chi,\Upsilon\}\cr
&& + e \{\Psi^\vee,\chi\} - i e [\Phi^\vee,f] - i e [\Phi^\vee,d\V\*\Phi_{-1}]
\eea
and now we shall expand 
\bea
D\h{\L_U}\Phi &=& [D,\h{\L_U}]\Phi + \h{\L_U}D\Phi\cr
&=& i e [\iota_U \F,\Phi] + d\V \* \h{\L_U} \Phi_{-1} + ...\cr
&\leadsto & d\V \* \h{\L_U} \Phi_{-1}
\eea
and 
\bea
D\h{\L_V}\Phi^\vee &=& [D,\h{\L_V}]\Phi^\vee + \h{\L_V}D\Phi^\vee\cr
&=& i e [\iota_V \F,\Phi^\vee] + ...\cr
&\leadsto & i e [f,\Phi^\vee]
\eea
Then
\bea
\delta (r.h.s) &=& \t{D\R}^+ - d\V \* \h{\L_U} \Phi_{-1} - i e [f,\Phi^\vee] + e \{\chi,\Upsilon\}\cr
&&  + e \{\Psi^\vee,\chi\} - i e [\Phi^\vee,f] - i e [\Phi^\vee,d\V\*\Phi_{-1}]
\eea
Now we note that the last commutator vanishes as one may see by the following rewriting,
\bea
[\Phi^\vee,d\V\*\Phi_{-1}] = - \U \times d\V \times [\Phi_{-1},\Phi_{-1}] = 0
\eea
By canceling some terms, we now find that the supersymmetry variation leads us to the selfdual projected Bianchi identity
\bea
\h{\L_U} f &=& - \t{D\R}^+
\eea
thus showing that this equation of motion is indeed supersymmetric.

\section{Five-dimensional super Yang-Mills}\label{tio}
We expect that any commutator terms in the five-dimensional super Yang-Mills theory will not be sensitive to the direction $n^M$ along which we perform the dimensional reduction. We do not want to put $n^M$ along neither $V^M$ nor along $U^M$ which are both lightlike directions. Instead we will in this section choose $n^M$ to be given by
\bea
n^M &=& V^M + U^M + \t{n}^M
\eea
which, if we assume $\N>0$, will be a spacelike conformal Killing vector field. Here we allow for a possibly nonvanishing transverse component $\t{n}^M$ in order to allow for a more general dimensional reduction when such a conformal Killing vector field $n^M$ exists on the given six-manifold. It is nontrivial task to perform dimensional reduction along a conformal Killing vector field. However, such a dimensional reduction was recently carried out in \cite{Gustavsson:2024yxh} along the conformal lightlike vector field $V$. While dimensional reduction along $n^M$ is inaccessible to us, we expect that as far as commutator terms in the resulting five-dimensional theory concern, those will not depend on the choice of conformal Killing vector field along which we reduce. 

The dimensional reduction along $V$ is implemented by imposing the constraints
\bea
\L_V \Psi_{MN} &=& - \frac{\Omega}{3} \Psi_{MN}\cr
\L_V \phi &=& - \frac{\Omega}{3} \phi
\eea
on the matter fields. Here we will not consider the gauge field since our goal will be to match the fermionic equations of motion in loop space with the corresponding fermionic equations of motion in the five-dimensional theory. We are particularly interested in the commutator terms here since for abelian tensor multiplet we already know that it reduces correctly to a corresponding five-dimensional vector multiplet. 
 
Using a Fierz identity in appendix \ref{spinors} we find that
\bea
\Psi^{aMN} \Psi^b_{MN} &=& - \frac{5}{4} \bar\lambda^{aI} \Gamma^P \lambda^b_I V_P - \frac{1}{4} \bar\lambda^{aI} \Gamma^{RST} \lambda^b_J \(\Theta_{RST}\)^J{}_I
\eea
and 
\bea
\psi^a \psi^b &=& - \frac{1}{8} \bar\lambda^{aI} \Gamma^M \lambda^b_I V_M + \frac{1}{24} \bar\lambda^{aI} \Gamma^{MNP} \lambda^b_J \(\Theta_{MNP}\)^J{}_I
\eea
from which it follows that
\bea
\bar\lambda^{aI} \Gamma^M \lambda^b_I V_M &=& - \frac{1}{2}  \Psi^{aMN} \Psi^b_{MN} - 3 \psi^a \psi^b
\eea
If we expand
\bea
\Psi^{aMN} \Psi^b_{MN} &=& \chi^{aMN} \chi^b_{MN} - 2 \psi^a \psi^b
\eea
then the result can also be written as
\bea
\bar\lambda^{aI} \Gamma^M \lambda^b_I V_M &=& - \frac{1}{2}  \chi^{aMN} \chi^b_{MN} - 2 \psi^a \psi^b
\eea
In the five-dimensional super Yang-Mills Lagrangian we have the fermionic terms
\bea
\L_F &=& \frac{i}{2} \bar\lambda^{aI} \Gamma^M D_M \lambda^b_I g_{ab} - \frac{i e}{2} \bar\lambda^{aI} \Gamma^M \lambda^{b}_I V_M  \phi^c f_{abc} 
\eea
Our Lie algebra conventions are 
\bea
[t_a,t_b] &=& i f_{ab}{}^c t_c\cr
\tr\(t_a t_b\) &=& g_{ab}\cr
f_{abc} &=& f_{ab}{}^d g_{dc}\cr
g^{ab}g_{bc} &=& \delta^a_c
\eea
To obtain the nonabelian Lagrangian in cohomological form, we will follow the following strategy. First we note that we need to replace $\nabla_M$ with $D_M$ in the correspondence
\bea
\frac{i}{2} \bar\lambda^{I} \Gamma^M \nabla_M \lambda_I &=& - \frac{i}{4\N} \chi^{MN} \L_U \chi_{MN} + \frac{i}{4\N} \chi^{MN} \partial_M \psi_N \cr
&& - \frac{i}{\N} V_{MN} \psi^M \psi^N\cr
&& + \frac{i}{2\N^2} \psi^M \L_V \psi_M - \frac{i}{\N^2} \psi^M \partial_M \(\N\psi\) - \frac{i}{\N^2} \psi \L_U \(\N \psi\)
\eea 
to get the corresponding nonabelian terms. Here we also notice that doing this in the Lie derivatives will map them into covariant Lie derivatives. Thus the nonabelian correspondence between spinor and tensor fermionic terms in the Lagrangian becomes 
\bea
\frac{i}{2} g_{ab} \bar\lambda^{aI} \Gamma^M D_M \lambda^b_I &=& - \frac{i}{4\N} g_{ab} \chi^{aMN} \h{\L_U} \chi^b_{MN} + \frac{i}{4\N} g_{ab} \chi^{aMN} D_M \psi^b_N \cr
&& - \frac{i}{\N} g_{ab} V_{MN} \psi^{aM} \psi^{bN}\cr
&& + \frac{i}{2\N^2} g_{ab} \psi^{aM} \h{\L_V} \psi^b_M - \frac{i}{\N^2} g_{ab} \psi^{aM} D_M \(\N\psi^b\)\cr
&& - \frac{i}{\N^2} g_{ab} \psi^a \h{\L_U} \(\N \psi^b\)
\eea
and
\bea
- \frac{i e}{2} \bar\lambda^{aI} \Gamma^M \lambda^{b}_I V_M  \phi^c f_{abc} &=& i e \(\frac{1}{4} \chi^{aMN} \chi^b_{MN} + \psi^a \psi^b\) \phi^c f_{abc}  
\eea
From this Lagrangian we then obtain the three fermionic equations of motion
\bea
\frac{\N^2}{\sqrt{-g}} D_M \(\frac{\sqrt{-g}}{\N^2} g^{MN} \psi^a_N\) + \frac{2}{\N} \h{\L_U} \(\N\psi^a\) &=& - 2 \N e \psi^b \phi^c f_{bc}{}^a\cr
\N^2 \t{D^M \(\frac{1}{\N} \chi^a_{MN}\)} + \h{\L_V} \psi^a_N &=& \t{D}_N \(\N\psi\)^a + \N V_{NP} \psi^{aP}\cr
\h{\L_U} \chi^a_{MN} &=& \t{\(D_M \psi^a_N - D_N \psi^a_M\)}^+ + \N e \chi^b_{MN} \phi^c f_{bc}{}^a \cr
&&\label{fivedimna}
\eea
by varying $\psi^a$, $\psi^a_M$ and $\chi^a_{MN}$ respectively. 

Let us pause here and look more closely at the equations that we have found. In the nonabelian five-dimensional theory we have got the equations of motion in (\ref{fivedimna}). These equations of motion are expressed in a form where six-dimensional covariance is manifest. The reduction to five-dimensions is implemented instead by imposing constraints on the fields on the form
\bea
\h{\L_V} \varphi &=& \frac{c_\varphi \Omega_V}{6} \varphi\label{cV}
\eea
Here $c_{\varphi}$ denotes the Weyl weight of the given field. Furthermore, if we compare the resulting equations of motion in (\ref{fivedimna}) with the corresponding abelian six-dimensional requation of motion (\ref{firsteom}), (\ref{secondeom}) and (\ref{thirdeom}) then we see that by putting the gauge group to be abelian in (\ref{fivedimna}) then those equations of motion exactly agree with (\ref{firsteom}), (\ref{secondeom}) and (\ref{thirdeom}). This leads us to make the following conjecture. We conjecture that the equations of motion we get upon dimensional reduction along $n^M$ are again given by (\ref{fivedimna}) but what changes now is only the constraints (\ref{cV}) that should be replaced with 
\bea
\h{\L_n} \varphi &=& \frac{c_\varphi \Omega_n}{6} \varphi\label{genn}
\eea

The last equation of motion resembles the loop space nonabelian equation of motion
\bea
\h{\L_U} \chi &=& \t{D \Upsilon}^+ + i e [\Phi^\vee,\chi]
\eea
We can make the correspondence precise by using our explicit realization
\bea
\h{\L_U} \chi &=& \int \h{\L_U} \chi_{MN}^a \delta C^M n^N t_a(s) ds\cr
\t{D \Upsilon}^+ &=& \int \(D_M \psi_N^a - D_N \psi_M^a\)^+ \delta C^M n^N t_a(s) ds\cr
i e [\Phi^\vee,\chi] &=& - e \int \phi^a \chi_{MN}^b U_P n^P \delta C^M n^N f_{ab}{}^c t_c(s) ds 
\eea
Thus we obtain a precise match if we take 
\bea
U_P n^P &=& \N
\eea
and assume that all local fields are constant along the loop under consideration, and if we define 
\bea
t_a &=& \int t_a(s) 
\eea

From the 5d equation of motion 
\bea
\frac{\N^2}{\sqrt{-g}} D_M \(\frac{\sqrt{-g}}{\N^2} g^{MN} \psi^a_N\) + \frac{2}{\N} \h{\L_U} \(\N\psi^a\) &=& - 2 \N e \psi^b \phi^c f_{bc}{}^a
\eea
we conjecture that we shall have the fermionic equation
\bea
D^\dag \Upsilon &=& 2 \h{\L_U} \Psi_{-1} - 2 i e [\Phi^\vee,\Psi_{-1}]
\eea
in loop space for nonabelian a gauge group. The appearance of the combination $\C + \Phi^\vee$ in this conjectured equation of motion is appealing given its close relation with $\Upsilon$ by supersymmetry, as
\bea
\delta \(\C + \Phi^\vee\) &=& i \Upsilon\cr
\delta \Upsilon &=& \L_U \B - \L_V \(\C + \Phi^\vee\) + i e [\B,\C + \Phi^\vee]
\eea
Under dimensional reduction along $n^M = V^M + U^M$ we shall restrict
\bea
\h{\L_n} \;=\; \h{\L_V} + \h{\L_U}
\eea
on all the fields, according to their Weyl weights. Together with $\iota_n \A = 0$ (from $A_M^a n^M = B_{MN}^a n^M n^N = 0$) this leads to 
\bea
[\h{\L_U},\h{\L_V}] &=& 0
\eea
on all fields, which amounts to putting  
\bea
\R &=& 0
\eea
for any semisimple nonabelian gauge group. This in turn implies that now the dimensional reduction of the fermionic equation of motion (\ref{xi}) results in that we shall put $\xi = 0$ since otherwise its supersymmetry variation will be nonvanishing contradicting the fact that $\R = 0$. Thus upon dimensional reduction we descend to
\bea
D^\dag\chi &=& \h{\L_V} \Upsilon - \t{D\Psi_{-1}} - * \(d\V \* \Upsilon\)
\eea
that now matches with the corresponding equation of motion in 5d that does not couple to the scalar field,
\bea
- \N^2 \t{D^M \(\frac{1}{\N} \chi_{MN}\)} &=& \h{\L_V} \psi_N - \t{D}_N \(\N\psi\) - \N V_{NP} \psi^P 
\eea
If we perform dimensional reduction along a more general conformal Killing vector $n^M$ not necessarily equal to $V^M + U^M$ we expect to get the same form of the equation of motion as written in a six-dimensional covariant form, with the only difference then being that we impose the restriction on all fields on the form (\ref{genn}) with this more general vector field $n^M$. Thus our strategy in this section has been to always derive the equations of motion by dimensional reduction along the most convenient choice of conformal Killing vector field and by having the equation in a six-dimensional form, we expect the same resulting equation of motion holds for any dimensional reduction. However, this strategy only works for the matter part of the Lagrangian. The gauge field part requires a careful separate analysis where we need separate treatments depending on whether we consider lightlike dimensional reduction or spacelike dimensional reduction.

\section{Discussion}\label{elva}
There are several inequivalent ways that one may introduce a nonabelian transgression map. Below we enumerate a few of these nonabelian transgression and transgression-like maps.
\bea
\P_1(\omega) &=& \int ds\omega^a_{M_1 \cdots M_p}(C(s)) \delta C^{M_1} \wedge \cdots \wedge \delta C^{M_{p-1}} \dot{C}^{M_p} t_a\cr
\P_2(\omega) &=& \int ds (W^{-1}\omega^a_{M_1 \cdots M_p}t_a W)(C(s)))\delta C^{M_1} \wedge \cdots \wedge \delta C^{M_{p-1}} \dot{C}^{M_p}\cr
\P_3(\omega) &=& \int ds\omega^a_{M_1 \cdots M_p}(C(s)) \delta C^{M_1} \wedge \cdots \wedge \delta C^{M_{p-1}} n^{M_p} t_a(s)\cr
\P_4(\omega) &=& \int ds\omega^a_{M_1 \cdots M_p}(C(s)) \delta C^{M_1} \wedge \cdots \wedge \delta C^{M_{p}} t_a(s)\cr
\P_5(\omega) &=& \sum_{i \in \text{Cartan}} \int ds\omega^i_{M_1 \cdots M_p}(C(s)) \delta C^{M_1} \wedge \cdots \wedge \delta C^{M_{p-1}} \dot{C}^{M_p} t_i\cr
&&+ \sum_{\alpha \in \text{roots}}\int ds\omega^\alpha_{M_1 \cdots M_p}(C(s)) \delta C^{M_1} \wedge \cdots \wedge \delta{C}^{M_p} t_\alpha(s)\cr
\P_6(\omega) &=& \int ds \omega_{M_1 \cdots M_p}(s,C) \delta C^{M_1} \wedge \cdots \wedge \delta C^{M_p} t_a(s)\cr
\P_7(\omega) &=& \int ds \omega_{M_1 \cdots M_p}(s,C) \delta C^{M_1} \wedge \cdots \wedge \delta C^{M_p} t_a\cr
\P_8(\omega) &=& ?
\eea
Let us describe shortly each of these maps:
\begin{enumerate}
\item
The map $\P_1$ does not immediately lead to a reparametrization invariant Wilson surface. But we like to caution that ruling out $\P_1$ completely might be premature since perhaps one can define a Wilson surface observable as a sum over all different parametrizations of a geometric surface. That sum may then become reparametrization invariant. The resulting field theory appears to become nonlocal. But the nonabelian tensor multiplet theory is expected to be local. 
\item
The search for a basic (as opposed to an average over parametrizations) reparametrization invariant Wilson surface has led to $\P_2$ that uses a Wilson line $W$ to parallel transport $\omega_{M_1\cdots M_p}$ to a common reference point on the loop. This enables us to define a reparametrization invariant nonabelian Wilson surface while introducing Wilson lines and a constraint of a vanishing fake curvature of a gauge field strength \cite{Schreiber:2005ff}, \cite{Baez:2004in}, \cite{Girelli:2003ev}. Wilson lines are associated to geometric lines to which we do not associate a physical observable in the abelian tensor multiplet. As lines are geometric objects on the spacetime manifold, they do not couple nontrivially to the gauge group bundle. So we do not expect line observables in the nonabelian tensor multiplet either. The vanishing fake curvature leads to a local field theory but the presence of line operators and a vanishing fake curvature condition has made it difficult to supersymmetrize.
\item
In the present paper we introduced $\P_3$ that makes use of a loop algebra and introduces a vector field $n$. This led to the requirement of dimensional reduction along $n$ and to the five-dimensional super-Yang-Mills description but now reformulated in a loop space language. Although we already knew that the five-dimensional super-Yang-Mills theory is one way to characterize the nonabelian tensor multiplet, the result is also showing that the loop algebra is an essential ingredient for an alternative way to describe the nonabelian tensor multiplet in loop space. 
\item
The map $\P_4$ is somewhat related to $\P_3$ as it is obtained by absorbing $n$ into a new form-field that we then again call as $\omega_{M_1\cdots M_p}$ as we describe the map $\P_4$. This way $\P_4$ becomes a rank-preserving map from $\Omega^p(M)$ to $\Omega^p(LM)$ for nonabelian $p$-forms. Again we use the loop algebra but now we do not introduce a vector field $n$ in this transgression map. This seems to lead to a 6d Yang-Mills theory. It is not clear to us whether this can be supersymmetrized. 
\item
The map $\P_5$ is a modification of $\P_4$ that blends in the abelian transgression for the Cartan generators. This may be useful for describing a Higgs mechanism with Cartans being massless tensor multiplet particles interacting with massive string fields valued in the loop algebra roots. The loop algebra valued fields are expected to be massive strings whose mass have been acquired via a Higgs mechanism. Again supersymmetrization is then just as unclear to us as it was for $\P_4$.
\item
The map $\P_6$ may be viewed as an expansion of the field on $LM$ in its loop algebra components. The map $\P_6$ includes $\P_3$ and $\P_4$ as special cases. 
\item
The map $\P_7$ might be another possibility that is valued in a ordinary Lie algebra instead. The map $\P_7$ includes $\P_1$ and $\P_2$ as special cases.
\item
We can not be sure if the list we presented above is exhaustive. We have included a map $\P_8$ that may be discovered in the future, perhaps by studying a discretization such as a lattice formulation.  
\end{enumerate}

The general strategy is to start from the abelian tensor multiplet, reformulating the spinorial fields in terms of tensor fields, then uplift these tensor fields to loop space via an abelian transgression map. We then seek a nonabelian generalization in loop space that we base on either $\P_6$, $\P_7$ or $\P_8$. That approach is natural because in loop space we will always have a one-form gauge field so we may construct a nonabelian generalization in loop space. If we base it on $\P_6$ then we may subsequently specialize to either $\P_3$ as we did in the present paper, or on $\P_4$ and $\P_5$ as we plan to do in future works. If instead we base it on $\P_7$ then we may subsequently specialize that to $\P_1$ or $\P_2$. For each choice of $\P_m$ ($m=1,...,8$) it may turn out that either a supersymmetric extension can not be found or if can be found the description may be gauge anomalous or if it is not gauge anomalous it may lead back to what we already knew such as five-dimensional super-Yang-Mills, or else it may give us a new, yet not known description of the theory. 

Let us now discuss $\P_4$ that leads to a seemingly ordinary Yang-Mills description on spacetime through the transgression map
\bea
\A(C) &=& \int ds A_M^a(C(s)) \delta C^M(s) t_a(s)
\eea
Let us compute its field strength in some detail here, as that computation illustrates how we use the loop algebra and the wedge product in loop space in a concrete computation. The field strength in loop space is always defined as
\bea
\F(C) &:=& \delta \A(C) - i e \A(C) \wedge \A(C)
\eea
and that is true regardless of what transgression map we use. Let us here only present how we shall expand the wedge product term if we use the map $\P_4$,
\ben
- i e \A(C) \wedge A(C) &=& - i e \int ds \int ds' A_M^a(C(s)) A_N^b(C(s')) t_a(s) t_b(s') \delta C^M(s) \wedge \delta C^N(s')
\een
Now because the wedge product on loop space is antisymmetric
\ben
\delta C^M(s) \wedge \delta C^N(s') &=& - \delta C^N(s') \wedge \delta C^M(s)
\een
we can also write the result as
\ben
- i e \A(C) \wedge A(C) &=& i e \int ds' \int ds A_N^b(C(s')) A_M^a(C(s)) t_b(s') t_a(s) \delta C^M(s) \wedge \delta C^N(s')
\een
where we also relabeled the dummy variables $s, a, M$ and $s', b, N$. By adding these results together, we find that 
\ben
- i e \A(C) \wedge \A(C) &=& - i e \int ds \int ds' A_M^a(C(s)) A_N^b(C(s')) [t_a(s),t_b(s')] \delta C^M(s) \wedge \delta C^N(s')
\een
Now we use the loop algebra and get a local expression integrated over the loop,
\bea
- i e \A(C) \wedge \A(C) &=& e f_{ab}{}^c \int ds A_M^a(C(s)) A_N^b(C(s)) t_c(s) \delta C^M(s) \wedge \delta C^N(s)
\eea
The final result we get is
\bea
\F(C) &=& \int ds F_{MN}^a(C(s)) t_a(s) \delta C^M(s) \wedge \delta C^N(s)
\eea
where 
\bea
F_{MN}^a &=& \partial_M A_N^a - \partial_N A_M^a + e  f_{bc}{}^a A_M^b A_N^c
\eea
We get the ordinary Yang-Mills field strength on the nose. In loop space it is natural to propose that we have the Yang-Mills term 
\bea
- \frac{1}{4} \tr(\F(C),\F(C))_{LM}
\eea
If we then use $\P_4$, then that would lead to a Yang-Mills term in spacetime, $- \frac{1}{4} \tr(F,F)_M$. From the way that we have defined the Weyl invariant inner product in (\ref{innerproductN}), that leads to the following Yang-Mills action in spacetime
\bea
S &=& - \frac{1}{4} \int_M d^6 x \sqrt{-g} g^{MP} g^{NQ} \frac{1}{\N} F^a_{MN} F^b_{PQ} g_{ab} \label{SYM}
\eea
where $g_{ab} = \tr(t_a t_b)$. This action is Weyl invariant because of the Weyl weight $\frac{1}{\N}$. That Weyl weight also makes the action reduce to the 5d Yang-Mills action with a 5d Yang-Mills coupling constant $g_{5d} \sim R$ where $R$ is the radius of the compactified circle.\footnote{If the 6d metric is related to the 5d metric as $ds^2_{6d} = ds^2_{5d} + R^2 dy^2$ where the 5d metric is $ds^2_5 = G_{\mu\nu} dx^{\mu} dx^{\nu}$ and $y\sim y + 1$ parametrizes the circle and if $\N \sim R^2$ then we get our 6d Yang-Mills action as $S_{6d} \sim \frac{1}{R} \int d^5 x \sqrt{-G} F_{\mu\nu}^2$. The key new ingredient to get this result is the weight factor $\frac{1}{\N} \sim \frac{1}{R^2}$. For this argument we have put the 6d coupling to be $e \sim 1$ as that is what we expect to see upon quantization. The coupling $e$ does not acquire its proper fixed value in classical field theory and supersymmetry alone is not enough to fix the value of $e$.} Our loop space techniques always generates Weyl invariant terms in the action on spacetime. Weyl invariance is guaranteed whenever we start with an action in loop space transgressed back to spacetime because our fields in loop space are all constructed to be Weyl invariant. The scalar field kinetic term in loop space that we expect to be of the form 
\bea
- \frac{1}{2} \tr(D\Phi,D\Phi)_{LM} &=& -\frac{1}{2} (D(\N\phi)^a,D(\N\phi)^b)_M g_{ab}
\eea
will, when expanded out, contain the correct conformal mass term and is guarenteed to be Weyl invariant since the inner product is constructed with a Weyl weight to make it Weyl invariant for a Weyl invariant one-form field such as here $D(\N\phi)^a$. We present the detailed computations that illustrates this for an abelian gauge group in appendix $D$. 

For the little Lorentz times R-symmetry group $Spin(4)_L \times SU(2)_R = SU(2) \times SU(2)\times SU(2)_R$ the vector multiplet has a vector particle in $(2,2;1)$ and a fermionic particle in $(1,2;2)$. Notably, the vector multiplet has no scalar field. For the same symmetry group the tensor multiplet has a selfdual particle in $(3,1;1)$, one scalar field in $(1,1;1)$ and fermions in $(2,1;2)$. From our procedure where we start from the abelian tensor multiplet, transgress, nonabelianize and then descend back via inverse nonabelian transgression, it is clear that we will still have one scalar field in our supermultiplet. It can not be that we by such a procedure we would go from the abelian tensor multiplet and end up with the nonabelian vector multiplet. That would be an impossibility.

The question then arises if this shall realize the nonabelian tensor multipet, how do we find the chiral representation $(3,1;1)$ from $F^a_{MN}$? Here we are not sure about the answer, the details needs to be worked out, if at all that will be possible. Let us simply note the following. Chirality in the abelian tensor multiplet is implemented on a 6d covariant tensor field in a 6d covariant way as $H_{MNP} = \frac{1}{6} \eps_{MNP}{}^{RST} H_{RST}$ and it leads to the representation $(3,1;1)$ of the little group there. For a two-form $F_{MN}$ in 6d there is no 6d covariant way to implement a selfdual condition, unless there is some additional geometric structure. What we noticed in \cite{Bak:2024ihe} is that for each choice of $(1,0)$ supersymmetry parameters that we use to define our theory in the cohomological formulation in which fermionic fields are $p$-form fields (a necessary precursor for a subsequent loop space reformulation), we have two commuting lightlike conformal Killing vector fields $V$ and $U$. Thus as we consider a 6d covariant theory equipped with $U$ and $V$ we have a covariant way to impose a selfduality condition on $F_{MN}$ by first projecting the components of $F_{MN}$ onto the transverse four-dimensional space orthogonal to both $V$ and $U$ and in that projected space we may impose selfduality on the projected components of $F_{MN}$. We do not know if this is how we may actually realize $(3,1;1)$ and let us also notice that the four dimensional space associated to the little group is not in general the same as this transverse space.  

We can not extend our Yang-Mills term to a full $(1,0)$ supersymmetric extension unless we use the vector multipet. What we may hope for is that if we select two supercharges from those at most $16$ supercharges on flat $\mb{R}^{1,5}$ (or some reduced amount of supercharges on some curved six-manifold) then we may extend the Yang-Mills action supersymmetrically while preserving only those two supercharges. For each choice we make of two supercharges we induce a different set of $U$ and $V$ vector fields and a different breaking of the Lorentz group. (In particular $V$ is the Dirac current of the supersymmetry parameter.) 

Leaving the question of a possible supersymmetric extension aside for the moment, then as far as the action (\ref{SYM}) concerns, the loop algebra does not appear, as only the component fields that are contracted with $g_{ab}$ of the Lie algebra of the gauge group appear. This obscures how we shall interpret the action, as to whether it shall be an action of particles or an action of tensionless strings. Our starting point was a loop algebra in loop space and in spacetime we may introduce a surface holonomy as
\bea
W &=& P \exp i e \int dt \int ds A^a_M(C(t,s)) \frac{\partial C^M(t,s)}{\partial t} t_a(s)\label{SH}
\eea
This motivates us to associate the component fields $A^a_M(x)$ with loops going through that point as $x = C(s)$ and to that loop associate loop algebra generators $t_a(s)$ rather than the Lie algebra generators $t_a = \int ds t_a(s)$. This might be how two seemingly contradictory properties of the theory might coexist -- the nonabelian tensor multiplet theory is expected to be a local theory and a theory of tensionless strings.

We may get a heavy closed string via the Higgs mechanism. To this end, we would like to apply the transgression map $\P_5$. For $SU(2)$ gauge group spontaneously broken to $U(1)$ by a Higgs vacuum expectation value, we then introduce the Cartan generator $t_3 = \int ds t_3(s)$ and the root generators $t_{\pm}(s)$. These generators satisfy the algebra
\bea
[t_3,t_+(s)] &=& t_+(s)\cr
[t_3,t_-(s)] &=& - t_-(s)\cr
[t_+(s),t_-(s)] &=& 2\delta(s-s') t_3
\eea
and one may check that these commutation relations satisfy the Jacobi identity, so this is a Lie algebra. We then define
\ben
\A(C) &=& \int ds \(B_{MN}(C(s)) \dot{C}^N(s) t_3 + A^+_M(C(s) t_+(s) + A^-_M(C(s)) t_-(s)\) \delta C^M(s)
\een
and expand the action around a scalar field vacuum expectation value $\phi_0$ as 
\ben
\Phi(C) &=& \int ds V_M(C(s)) \(\phi_0^3(C(s)) + v(C(s))\) \dot{C}^M(s) t_3\cr
&& + \int ds \N \(\phi^+(C(s)) t_+(s) + \phi^-(C(s)) t_-(s)\)
\een
This way we might get a massless tensor field $B_{MN}$ that interacts with heavy modes induced by the Higgs mechanism. 

What we have discussed here are speculations and more research is needed regarding a possible supersymmetric extension.

\subsection*{Acknowledgement}
DB was supported in part by NRF Grant RS-2023-00208011, and by Basic Science Research Program through NRF Grant 2018R1A6A1A06024977 funded by the Ministry of Education. We like to thank the referee for suggesting improvements of especially section 3.

\appendix
\section{Our spinor conventions}\label{spinors}
We have the Fierz identity
\bea
\eps_I \bar\eps^J &=& \frac{1}{8} \delta_I^J \Gamma_M V^M - \frac{1}{24} \Gamma_{MNP} \(\Theta^{MNP}\)^J{}_I
\eea
where $V_M = \bar\eps^I \Gamma_M \eps_I$ and $\(\Theta_{MNP}\)^I{}_J = \bar\eps^I \Gamma_{MNP} \eps_J$. We raise and lower indices as $\psi^I \epsilon_{IJ} = \psi_J$ and $\psi^I = \epsilon^{IJ} \psi_J$ where $\epsilon_{IJ} \epsilon^{JK} = - \delta_I^K$. Six-dimensional gamma matrices are denoted $\Gamma_M$ that obey
\bea
\{\Gamma_M,\Gamma_N\} &=& 2 g_{MN}
\eea
The charge conjugation matrix in six dimensions is denoted $C$ that has the properties
\bea
C^T &=& C\cr
\Gamma_M^T &=& - C \Gamma_M C^{-1}
\eea
We have the symplectic-Majorana condition
\bea
\bar\eps^I &=& \epsilon^{IJ} \eps_J^T C\cr
\bar\lambda^I &=& \epsilon^{IJ} \lambda_J^T C
\eea
where the Dirac conjugates are defined as $\bar\eps^I = (\eps_I)^* \Gamma^{\hat{0}}$ and $\bar\lambda^I = (\lambda_I)^* \Gamma^{\hat{0}}$ that uses flat tangent space gamma-zero, so that $\Gamma^{\hat{0}} \Gamma^{\hat{0}} = - 1$. By introducing the inverse $\t\epsilon^{IJ} = - \epsilon^{IJ}$ satisfying $\epsilon_{IJ} \t\epsilon^{JK} = + \delta_I^K$ then we can write the symplectic Majorana condition in the form
\bea
\bar\eps^I &=& \eps_J^T \t\epsilon^{JI} C
\eea
We have (assuming $\eps_J$ is commuting)
\bea
\Psi_{MN} &=& \bar\eps^I \Gamma_{MN} \lambda_I\cr
&=& \eps_J^T \t\epsilon^{JI} C \Gamma_{MN} \lambda_I\cr
&=& \(\eps_J^T \t\eps^{JI} C \Gamma_{MN} \lambda_I\)^T\cr
&=& - \lambda_I^T C \Gamma_{MN} \eps_J \t\eps^{JI}\cr
&=& \bar\lambda^I \Gamma_{MN} \eps_I
\eea
The transpose on $\eps_I^T$ does not do anything, but we write it in order to better see how it is being contracted with matrices on the right of it and also in order to better see how transpose acts. If we write out the chiral spinor indices $A,B,... = 1,...,4$ downstairs as in $\lambda_{AI}$ and antichiral spinor indices $A,B,...=1,...,4$ upstairs as in $\eps^A_I$, then we have six-dimensional gamma matrices on the form
\bea
\(\begin{matrix}
0 && \(\Gamma^M\)_{AB}\\
\(\Gamma^M\)^{AB} && 0
\end{matrix}\)
\eea
the six-dimensinoal charge conjugation matrix
\bea
C &=& \(\begin{matrix}
C_A{}^B && 0\\
0 && C^A{}_B
\end{matrix}\)
\eea
and the fermionic tensor field
\bea
\Psi_{MN} &=& \eps_J^A \t\epsilon^{JI} C_A{}^B \(\Gamma_{MN}\)_B{}^C \lambda_{CI}
\eea
Since the charge conjugation does not map a chiral spinor into an antichiral spinor we can distinguish them by their positions upstairs and downstairs, i.e. we do not need to use dotted indices for antichiral spinor indices. We have 
\bea
\(C\Gamma_M\)_{AB} &=& - \(C \Gamma_M\)_{BA}\cr
\(C\Gamma_{MN}\)_A{}^B &=& - \(C \Gamma_{MN}\)^B{}_A\cr
C_A{}^B &=& C^B{}_A
\eea
and 
\bea
\Psi_{MN} &=& \eps_J^A \t\epsilon^{JI} \(C\Gamma_{MN}\)_A{}^B \lambda_{BI}\cr
&=& \lambda_{BI} \t\eps^{IJ} \(\Gamma_{MN}\)^B{}_A \eps_J^A\cr
&=& \bar\lambda^I \Gamma_{MN} \eps_I
\eea
We also define a fermionic scalar field as
\bea
\psi &=& \bar\eps^I \lambda_I\cr
&=& \eps_J^T \t\epsilon^{JI} C \lambda_I\cr
&=& \lambda_I^T C \eps_J \t\epsilon^{JI}\cr
&=& - \bar\lambda^I \eps_I
\eea
or if we write the spinor indices
\bea
\psi &=& \eps_J^A \t\epsilon^{JI} C_A{}^B \lambda_{IB}\cr
&=& - \lambda_{IA} \t\epsilon^{IJ} C^B{}_A \eps_J^A
\eea

\section{Some properties of $U_{MN}$ and $V_{MN}$}\label{UoV}
In this appendix we will prove that $V_{MN}$ and $U_{MN}$ live in the transverse space where they are selfdual and antiselfdual respectively. 

From 
\bea
\Gamma_M \eps V^M &=& 0
\eea
we get
\bea
0 = \Gamma_N \Gamma_M \eps V^M U^N = \N \eps - \Gamma_{MN} \eps V^M U^N
\eea
So we have the Weyl projection
\bea
\Gamma_{MN} \eps \frac{V^M U^N}{\N} &=& \eps\label{2d}
\eea
We decompose 
\bea
\Gamma_M &=& \t\Gamma_M + \frac{1}{\N} V_M U^N \Gamma_N + \frac{1}{\N} U_M V^N \Gamma_N
\eea
Then 
\bea
V_{MN} := \nabla_M \(\frac{V_N}{\N}\) - \nabla_N \(\frac{V_M}{\N}\) = \frac{1}{\N} \(\nabla_M V_N - \nabla_N V_M\) - \frac{\nabla_M \N V_N - \nabla_N \N V_M}{\N^2}
\eea
On the other hand, we know that $V_{MN}$ is traceless because
\bea
V^M V_{MN} &=& \frac{1}{\N} V^M \nabla_M V_N - \frac{1}{\N^2} V_N V^M \nabla_M \N \cr
&=& \frac{\Omega}{3\N} V_N - \frac{1}{\N^2} V_N \L_V \N = 0\cr
U^M V_{MN} &=& \frac{1}{\N} U^M \(\nabla_M V_N - \nabla_N V_M\) - \frac{1}{\N^2} V_N \L_U \N + \frac{1}{\N} \nabla_N \N\cr
&=& \frac{1}{\N} U^M \(\nabla_M V_N - \nabla_N V_M\) - \frac{\Omega^\vee}{3\N} V_N + \frac{1}{\N} \nabla_N\N\cr
&=& \frac{1}{\N} \(U^M \nabla_M V_N - V^M \nabla_M U_N\) = 0
\eea
where we have used
\bea
\L_U \N &=& \frac{\Omega^\vee}{3} \N\cr
\frac{1}{\N}\nabla_N \N &=& \frac{\Omega^\vee}{3\N} V_N - \frac{1}{\N} V^M \nabla_M U_N + \frac{1}{\N} U^M \nabla_N V_M
\eea
 so we must have
\bea
V_{MN} &=& \frac{1}{\N} \(\t{\nabla_M V_N} - \t{\nabla_N V_M}\)
\eea
where the tilde means the following concrete thing. First we compute the thing under the tilde without the tilde being there. We just remove it. Then we remove from the quantity that we got all the components that are sticking out in the directions spanned by $U$ and $V$ in all tensor indices. So this we can go ahead and compute. First the thing without the tilde is obviously given by
\bea
\frac{1}{\N} \({\nabla_M V_N} - {\nabla_N V_M}\) &=& - \frac{4}{\N} \eps \Gamma_{MN} \eta
\eea
Then removing the things sticking out is easy, we just put tildes on the gamma matrices. So we end up with the result
\bea
V_{MN} &=& - \frac{4}{\N} \eps \t\Gamma_{MN} \eta
\eea
We now use the 6d Weyl condition of the form 
\bea
\Gamma_{MNPQRS} \eps &=& - \eps_{MNPQRS} \eps
\eea
that we may contract with $V^R U^S/\N$. If we also define 
\bea
\E_{MNPQ} &=& \eps_{MNPQRS} \frac{V^R U^S}{\N}
\eea
then the 6d Weyl condition becomes
\bea
\t\Gamma_{MNPQ} \eps &=& -\E_{MNPR} \eps
\eea
where we have also used the 2d Weyl condition (\ref{2d}). Thus by combining the 6d Weyl condition with the 2d Weyl condition, we have obtained a 4d Weyl condition. We now use this 4d Weyl condition to demonstrate that $V_{MN}$ is selfdual. We have
\bea
\bar\eps \t\Gamma_{MN} \eps = - \frac{1}{2} \bar\eta \t\Gamma^{PQ} \t\Gamma_{MNPQ} \eps = \frac{1}{2} \E_{MNPQ} \bar\eta \t\Gamma^{PQ} \eps
\eea
Thus we have 
\bea
V_{MN} &=& \frac{1}{2} \E_{MNPQ} V^{PQ}
\eea
A similar argument can be used to show that 
\bea
U_{MN} &=& - \frac{1}{2} \E_{MNPQ} U^{PQ}
\eea
where we define 
\bea
U_{MN} &=& \partial_M \frac{U_N}{\N} - \partial_N \frac{U_M}{\N}
\eea
This in particular implies that 
\bea
V^{MN} U_{MN} &=& 0
\eea

\section{Conformal Killing vectors}
We have two commuting lightlike conformal Killing vectors $U$ and $V$. Specifically
\bea
\L_V g_{MN} &=& \frac{\Omega}{3} g_{MN}
\eea
We decompose the metric as
\bea
g_{MN} &=& G_{MN} + \frac{1}{\N} \(U_M V_N + U_N V_M\)
\eea
By using $\L_V V^M = 0$ and $\L_V \frac{U^M}{\N} = 0$ we get 
\bea
\L_V \(\frac{1}{\N} U_M V_N\) &=& \frac{\Omega}{3} \frac{1}{\N} U_M V_N
\eea
and therefore we must have 
\bea
\L_V G_{MN} &=& \frac{\Omega}{3} G_{MN}
\eea
We then have
\bea
\L_V g &=& g g^{MN} \L_V g_{MN}\cr
&=& 2 \Omega g
\eea
and 
\bea
\L_V G &=& G G^{MN} \L_V G_{MN}\cr
&=& \frac{4\Omega}{3} G
\eea
We then have
\bea
\L_V \sqrt{-g} &=& \Omega \sqrt{-g}\cr
\L_V \sqrt{G} &=& \frac{2\Omega}{3} \sqrt{G}
\eea
If we now make the ansatz 
\bea
\sqrt{-g} &=& \alpha \sqrt{G}
\eea
then we conclude that $\alpha$ shall be such that 
\bea
\L_V \alpha &=& \frac{\Omega}{3} \alpha\cr
\L_U \alpha &=& \frac{\Omega}{3} \beta
\eea
Indeed there is a good candidate for $\alpha$ that satisfies these two requirements. It is $\alpha = \N$ up to some constant factor. We can always arrange it so that this factor is equal to one by a constant rescaling of $V^M$. Then we have the relation
\bea
\sqrt{-g} &=& \N \sqrt{G}
\eea

\section{A formula for the curvature scalar}\label{scalarR}
Let us begin by reviewing the story for a conformally flat six-manifold and how we may present its curvature scalar. We then have
\bea
g_{MN} &=& e^{2\sigma} \eta_{MN}\cr
e^A_M &=& e^{\sigma} \delta^A_M\cr
e^{AM} &=& e^{-\sigma} \eta^{MN} \delta^A_N\cr
e^M_A &=& e^{-\sigma} \delta_A^M\cr
\partial_M e^A_N &=& \partial_M \sigma e^A_N\cr 
\omega^{AB}_M &=& \(e^A_M e^{BN} - e^B_M e^{AN}\) \partial_N \sigma\cr
\partial_M e_N^A + \eta_{BC} \omega^{AB}_M e^C_N - \Gamma_{MN}^P e_P^A &=& \nabla_M e_N^A\cr
&=& 0
\eea
Covariant derivative
\bea
\nabla_M &=& \partial_M + \omega_M - \Gamma_M
\eea
Curvature
\bea
R^A{}_B &=& d \omega^A{}_B + c \omega^A{}_C \wedge \omega^C{}_B
\eea
We determine $c$ by demanding the Bianchi identity $\nabla R^{AB} = 0$ holds, or
\bea
d R^{AB} + \omega^A{}_C \wedge R^{CB} + \omega^B{}_C \wedge R^{AC} &=& 0
\eea
We get
\bea
d R^{AB} &=& c R^A{}_C \wedge \omega^C{}_B - c \omega^A{}_C \wedge R^C{}_B
\eea 
and because $R^{AB}$ is a two-form the wedge product is symmetric, so 
\bea
d R^{AB} &=& - c \omega^B{}_C \wedge R^{AC} - c \omega^A{}_C \wedge R^{CB}
\eea 
so we shall take $c= 1$ and then we have 
\bea
R^A{}_B &=& d \omega^A{}_B + \omega^A{}_C \wedge \omega^C{}_B
\eea
We then get
\bea
R &=& e^M_A e^N_B R_{MN}{}^{AB}\cr
&=& - 10 g^{MN} \(\partial_M \partial_N \sigma + 2 \partial_M \sigma \partial_N \sigma\)
\eea
The appearance of ordinary derivatives here may seem confusing. But the right way to look at these are as covariant derivatives with respect to the metric before the Weyl transformation took place, which was the flat Minkowski metric. So ordinary derivatives are correct to use here since they are indeed covariant with respect to $\eta_{MN}$.

But now we want to express this covariantly with respect to the new metric $g_{MN} = e^{2\sigma} \eta_{MN}$. We begin by expanding 
\bea
\nabla_M \partial_N \sigma &=& \partial_M \partial_N \sigma + \Gamma_{MN}^P \partial_P \sigma
\eea
where 
\bea
\Gamma^P_{MN} &=& \frac{1}{2} g^{PQ} \(-\partial_Q g_{MN} + \partial_M g_{NQ} + \partial_N g_{MQ}\)\cr
&=& \eta^{PQ} \(- \partial_Q \sigma \eta_{MN} + \partial_M \sigma \eta_{NQ} + \partial_N \sigma \eta_{MQ}\)\cr
&=& - g_{MN} \nabla^P \sigma + \delta^P_N \partial_M \sigma + \delta^P_M \partial_N \sigma
\eea
Then 
\bea
\nabla_M \partial_N \sigma &=& \partial_M \partial_N \sigma - g_{MN} \nabla^P \sigma \partial_P \sigma + 2 \partial_M \sigma \partial_N \sigma
\eea
and then 
\bea
g^{MN} \partial_M \partial_N \sigma &=& \nabla^2 \sigma + 4 \(\nabla_M\sigma\)^2
\eea
where on the right-hand side everything is written covariantly with respect to $g_{MN}$. Then 
\bea
R &=& - 10 g^{MN} \(\partial_M \partial_N \sigma + 2 \partial_M \sigma \partial_N \sigma\)\cr
&=& - 10 \nabla^2 \sigma + 40 \(\nabla_M \sigma\)^2 - 20 \(\nabla_M \sigma\)^2\cr 
&=& - 10 \nabla^2 \sigma + 20 \(\nabla_M \sigma\)^2 
\eea
and for $\sigma = \frac{1}{2} \log \N$ this becomes
\bea
\frac{1}{5} R &=& 2 \(\frac{\nabla_M \N}{\N}\)^2 - \frac{\nabla^2 \N}{\N}
\eea
For the metric $g_{MN} = \N \eta_{MN}$ and $\sqrt{-g} = \N^3$ the conformally invariant action becomes
\bea
S &=& - \frac{1}{2} \int d^6 x \sqrt{-g} \(g^{MN} \partial_M \phi \partial_N \phi + \frac{R}{5} \phi^2\)\cr
&=& \int d^6 x \sqrt{-g} \(-\frac{1}{2} g^{MN} \partial_M \phi \partial_N \phi - \frac{1}{2}\(2\(\frac{\nabla_M \N}{\N}\)^2 - \frac{\nabla^2 \N}{\N}\)\phi^2\)
\eea
On the other hand, the action that we are proposing is given by
\bea
S &=& - \frac{1}{2} \int d^6 x \sqrt{-g} g^{MN} \frac{1}{\N^2} \partial_M \(\N \phi\) \partial_N \(\N\phi\)\cr
&=& \int d^6 x \sqrt{-g} \(- \frac{1}{2} \(\nabla_M\phi\)^2 + \frac{1}{2} \(\frac{\nabla^2 \N}{\N} - 2\(\frac{\nabla_M \N}{\N}\)^2\) \phi^2\)
\eea
We see that we have agreement, at least for conformally flat spacetimes. 

But this agreement should be much more general. For the class of six-manifolds that we consider, we should always have
\bea
R &=& 10 \(\frac{\nabla_M \N}{\N}\)^2 - 5 \frac{\nabla^2 \N}{\N}\label{curvscalar}
\eea
irrespectively of whether the spacetime is conformally flat or not. Instead,  we are requiring the existence of two commuting conformal lightlike Killing vectors $U^M$ and $V^M$ where we define $\N = g_{MN} U^M V^N$. From this the result (\ref{curvscalar}) should follow from that the conformally invariant and diffeomorphism invariant action for a free scalar field is unique. As we have presented two such actions, by requiring these two actions to be identical, the relation (\ref{curvscalar}) should follow.

\section{Computation of $\nabla^M \(\frac{1}{\N} V_{MN}\)$}
We have
\bea
V_{MN} &=& - \frac{4}{\N} \bar\eps^I \t\Gamma_{MN} \eta_I
\eea
We have the Weyl projections
\bea
\Gamma_{MN} \eps_I  \frac{V^M U^N}{\N} &=& \eps_I\cr
\Gamma_{MNPQRS} \eps_I &=& - \eps_I \eps_{MNPQRS} 
\eea
from which it follows that 
\bea
\Gamma_{MNPQRS} \eps_I \frac{V^R U^S}{\N} &=& - \eps_I \eps_{MNPQRS} \frac{V^R U^S}{\N}\label{ida}
\eea
Now we notice that the antisymmetric product of gamma matrices factorizes
\bea
\t\Gamma_{MNPQ} \Gamma_{RS} \eps_I \frac{V^R U^S}{\N} &=& - \eps_I \eps_{MNPQRS} \frac{V^R U^S}{\N}
\eea
To see this, we may contract (\ref{ida}) with either $V^M$ or $U^M$ and see that both sides vanish. Now we can apply the Weyl projection above to get
\bea
\t\Gamma_{MNPQ} \eps_I &=& - \eps_I \E_{MNPQ} 
\eea
where we define 
\bea
\E_{MNPQ} &:=& \eps_{MNPQRS} \frac{V^R U^S}{\N}
\eea
Contracting with $\t\Gamma^{MN}$ gives
\bea
\t\Gamma_{PQ} \eps_I &=& \frac{1}{2} \t\Gamma^{MN} \eps_I \E_{MNPQ}
\eea
Writing
\bea
V_{MN} &=& - \frac{4}{\N} \bar\eta^I \t\Gamma_{MN} \eps_I
\eea
it is now clear that we have 
\bea
V_{MN} &=& \frac{1}{2} \E_{MNPQ} V^{PQ}
\eea
We have
\bea
\E_{MNPQ} \E^{MNPQ} &=& \eps_{MNPQRS} \eps^{MNPQUV} \frac{1}{\N^2} V^R U^S V_U U_V\cr
&=& - 4! 2! \delta^{UV}_{RS} \frac{1}{\N^2} V^R U^S V_U U_V\cr
&=& - 24 
\eea
and from 
\bea
\nabla_M \eps^{MNPQRS} &=& \frac{1}{\sqrt{-g}} \partial_M \(\sqrt{-g} \eps^{MNPQRS}\)\cr
&=& 0
\eea
since $\sqrt{-g} \eps^{MNPQRS} = \pm 1$, it follows that 
\bea
\nabla_M \(\frac{1}{\N} \E^{MNPQ}\) &=& \nabla_M \(\eps^{MNPQRS} \frac{U_R}{\N} \frac{V_S}{\N}\)\cr
&=& \frac{1}{2\N} \eps^{MNPQRS} \(U_{MR} V_S - V_{MR} U_S\)
\eea
This implies that 
\bea
\nabla_M \(\frac{1}{\N} \E^{MNPQ}\) &=& - \frac{3}{\N} \(U^{[Q} V^{NP]} + V^{[Q} U^{NP]}\)\label{derivedE}
\eea
which in turn implies that 
\bea
\nabla_M \(\frac{1}{\N} \E^{MNPQ} V_{PQ}\) &=& - \frac{1}{\N} U^N V^{PQ} V_{PQ}
\eea
Here we used the Bianchi identity
\bea
\nabla_{[M} V_{PQ]} &=& 0
\eea
and orthogonality constraints. Finally using selfduality we get
\bea
\nabla_M \(\frac{1}{\N} V^{MN}\) &=& - \frac{1}{2\N} U^N V^{RS} V_{RS}
\eea
To check the consistency of this result, we may contract both sides with $\frac{V_N}{\N}$. Then for the left-hand side
\bea
\frac{V_N}{\N} \nabla_M \(\frac{1}{\N} V^{MN}\) &=& \nabla_M \(\frac{V_N}{\N^2} V^{MN}\) - \nabla_M \(\frac{V_N}{\N}\) \frac{1}{\N} V^{MN}\cr
&=& - \frac{1}{2\N} V_{MN} V^{MN} 
\eea
and for the right-hand side,
\bea
- \frac{1}{2\N} V_{MN} V^{MN}
\eea
so they agree. Using selfduality, the result can be also expressed as
\bea
\nabla_M \(\frac{1}{\N} V^{MN}\) &=& - \frac{U^N}{4\N} \E^{RSPQ}V_{RS} V_{PQ}
\eea

\section{The fermionic selfduality equation}\label{FSD}
If we start with the bosonic selfduality equation of motion 
\bea
3\partial_{[M} B_{NP]} &=& \frac{1}{2} \eps_{MNP}{}^{RST} \partial_R B_{ST}
\eea
and make a supersymmetry variation, then we get a manifestly Lorentz invariant selfduality equation of motion for the fermionic field,
\bea
3\partial_{[M} \Psi_{NP]} &=& \frac{1}{2} \eps_{MNP}{}^{RST} \partial_R \Psi_{ST}
\eea
We will now show that from this equation of motion we reproduce the equation of motion (\ref{secondeom}) by contracting both sides with $V_M U_N$. When we contract the left-hand side, we get
\bea
3 V^M U^N \partial_{[M} \Psi_{NP]} &=& \L_V \(U^N \Psi_{NP}\) - \L_U \(V^N \Psi_{NP}\) - \partial_P \(V^M U^N \Psi_{MN}\)\cr
&=& \L_V \psi_N + \L_V \(U_N \psi\) + \L_U \(V_P \psi\) - \partial_P \(\N\psi\)\cr
&=& \L_V \psi_N - \t{\partial_P \(\N\psi\)}
\eea
Now let us turn to the right-hand side,
\bea
&& \frac{1}{2} V^M U^N \eps_{MNP}{}^{RST} \partial_R \Psi_{ST}\cr
&=& \frac{1}{2} V^M U^N \eps_{MNP}{}^{RST} \(\partial_R \chi_{ST} + 2\frac{V_M}{\N} \psi_N + ...\)\cr
&=& \frac{1}{2} V^M U^N \eps_{MNP}{}^{RST} \partial_R \chi_{ST} + \N V_{NP} \psi^P
\eea
Let us now focus on the first term
\bea
&& \frac{1}{2} V^M U^N \eps_{MNP}{}^{RST} \partial_R \chi_{ST}\cr
&=& \frac{\N^2}{2} \frac{V^M U^N}{\N^2} \eps_{MNP}{}^{RST} \nabla_R \chi_{ST}\cr
&=& \frac{\N^2}{2} \nabla_R \(\frac{V^M U^N}{\N^2} \eps_{MNP}{}^{RST} \chi_{ST}\) + ...\cr
&=& \frac{\N^2}{2} \nabla^R \(\frac{1}{\N} \E_{PR}{}^{ST} \chi_{ST}\) + ...
\eea
We now use off-shell selfduality 
\bea
\E_{PR}{}^{ST} \chi_{ST} &=& 2 \chi_{PR}
\eea
and get
\bea
&=& \N^2 \nabla^R \(\frac{1}{\N} \chi_{PR}\) + ...
\eea
Since the dots are terms with legs in the longitudinal directions of either $U^M$ or $V^M$, it is obvious, without any need for more detailed computations, that those terms will compensate for any such longitudinal components in $\N^2 \nabla^R \(\frac{1}{\N} \chi_{PR}\)$ so that the sum will be transverse, the longitudinal components have to be canceling out since the expression that we started with was transverse with no longitudinal components. We have therefore now derived the second equation of motion (\ref{secondeom})
\bea
\L_V \psi_N - \t{\partial_P \(\N\psi\)} &=& - \N^2 \nabla^M \(\frac{1}{\N} \chi_{MN}\) - \frac{\N}{2} U_N V^{PQ} \chi_{PQ} + \N V_{NP} \psi^P
\eea
from the manifestly Lorentz invariant fermionic selfduality equation of motion. 

In a similar way we can derive the equation of motion 
\bea
\L_U \chi_{MN} &=& \(\partial_M \psi_N - \partial_N \psi_M\)^+
\eea
by contracting the selfduality equation with $U^P$ on both sides and a subsequent projection onto the transverse space.

\end{document}